\documentclass[aps,amsmath,amssymb,prb,superscriptaddress, preprint]{revtex4-1}

\usepackage{verbatim}
\pdfoutput=1
\usepackage{dcolumn}
\usepackage{graphicx}
\usepackage{epstopdf}
\usepackage{amsfonts}
\usepackage{amsmath}
\usepackage{amssymb}
\usepackage{subfigure}
\usepackage{amssymb}
\usepackage{lipsum}
\usepackage{xcolor}
\usepackage{rotating}
\usepackage{wasysym}

\usepackage{booktabs}
\usepackage[colorlinks,bookmarks=false,citecolor=blue,linkcolor=red,urlcolor=blue]{hyperref}
\allowdisplaybreaks

\begin{document}
\title{Electron correlation effects in diamond: a wave-function quantum chemistry study of the quasiparticle band structure}

\author{Alexandrina\ Stoyanova}
\email[Corresponding authors: ]{alex07@mpipks-dresden.mpg.de;}
\affiliation{Max-Planck-Institut f\"ur Physik komplexer Systeme,
             N\"othnitzer Str.~38, 01187 Dresden, Germany}     
                                               
\author{Alexander\ O.\ Mitrushchenkov}
\affiliation{Laboratoire de Mod\'elisation et Simulation Multi Echelle, \\Equipe de Chimie Th\'eorique, 
5, Boulevard Descartes,
77454, Marne-la-Vall\'ee Cedex 2, France 
}     

\author{Liviu Hozoi} 
\affiliation{Institute for Theoretical Solid State Physics, IFW Dresden, Helmholtzstr.~20, 01069 Dresden, Germany} 
\author{Hermann Stoll}                                           
\affiliation{Institut f\"ur Theoretische Chemie, Universit\"at Stuttgart, Pfaffenwaldring 55, D-70569 
Stuttgart, Germany                 
               }
\author{Peter\ Fulde$^{1}$}

% ABSTRACT
\begin{abstract}
The quasiparticle bands of diamond, a prototype covalent insulator, are herein studied by means
of wave-function electronic-structure theory, with emphasis on the nature of the correlation hole around a bare particle.
Short-range correlations are in such a system conveniently described by using a real-space
representation and many-body techniques from {\it ab initio} quantum chemistry.
To account for long-range polarization effects, on the other hand, we adopt the approximation of a dielectric
continuum.
Having as ``uncorrelated'' reference the Hartree-Fock band structure,
the post-Hartree-Fock treatment is carried out in terms of localized Wannier functions derived from the Hartree-Fock solution.
The computed correlation-induced corrections to the relevant real-space matrix elements are important and give rise to a strong reduction, in the range of $50\%$, of the initial Hartree-Fock gap. While our final results for the indirect and direct gaps, 5.4 and 6.9 eV, respectively, compare very well with the experimental data, the width of the valence band comes out by $10$ to $15\%$ too large as compared to experiment.
This overestimation of the valence-band width appears to be related to size-consistency effects in the configuration-interaction correlation treatment.
\end{abstract} 
\maketitle

%%%%%%%%%%%%%%%INTRODUCTION%%%%%%%%%%%%%%%%%%%%%%
\section{Introduction}

Finding the optimal approach to accurate computations of energy bands in crystals still is
an active research topic in condensed matter theory.
Over the last few decades, the field has been dominated by formalisms based on density functional theory (DFT).
A central object within the DFT framework is the exchange-correlation functional and the most common approximations to it are the local density approximation (LDA) \cite {Perdew} and the generalized gradient approximation (GGA) \cite{Perdew2}. While formally a ground-state theory, DFT has been successfully applied to the computation of energy bands for a variety of systems. For metals, in particular, it often provides good agreement with $k$-space dispersions and Fermi surfaces determined by photoemission measurements.
The success has been limited, on the other hand, for semiconducting and insulating compounds
since with either the LDA or the GGA band gaps are systematically underestimated. This is typically referred to as the band-gap problem.
Such difficulties in canonical DFT are obviously not surprising because the correlation hole of an electron added to the conduction-band states or removed from the valence bands is  related to polarization effects, charge relaxation, and quantum fluctuations, and it is therefore different from that of electrons in the charge neutral ground state (GS).
In distinction to GS correlations, mostly of van der Waals type \cite{Fuldebook, Fulde, Fuldebook2},
the correlation hole of the additional particle in the $(N\!\pm\!1)$ configuration involves a substantially long-ranged (LR) polarization cloud.
The corresponding quasiparticle (QP), i.e., the bare particle plus its correlation hole, is moving in the form of a Bloch wave through the system. An accurate description of the energy bands requires an accurate description of the QP's, i.e., the correlation holes. The LDA and GGA potentials, for example, that depend only on the local electron density and gradient, respectively, are not able to correctly describe this correlation hole of the extra particle introduced in the system.
Improved results for band widths and band gaps can be obtained by post-LDA/GGA computational schemes~\cite{HL, HL2} based on the \emph{GW} approximation \cite{Hedin, Lundqvist}. The \emph{GW} method brings in the LR part of the correlation hole by including screening of the bare Coulomb interactions. It provides band gaps in reasonably good agreement with the experiment, see, e.g., Refs.~\onlinecite{Falco, Onida, Godby}. However, a treatment of short-range (SR) correlations, in particular, on-site correlations such as those in \emph{d}- and \emph{f}-metal oxides, is much more difficult within the \emph{GW} approach.

Other DFT-based formalisms aim at improving the approximate exchange-correlation functionals. Exact-exchange (EXX) treatments 
in Kohn-Sham (KS) band-structure schemes were introduced by Kotani~\cite{Kotani}, G\"orling ~\cite{Goerling}, and St\"adele \emph{et al.} \cite{Staedele1,Staedele2}, bringing semiconductor band gaps in very good agreement with experimental values~\cite{Staedele1,Staedele2, Rinke}.
Such EXX schemes gave rise to orbital-dependent functionals and to optimized effective potential (OEP) techniques, see, e.g., Refs.~\onlinecite{Kuemel, OEP}. Better agreement with the experiment for band gaps can be also achieved by using hybrid (see, e.g., Ref.~\onlinecite{Janesco}) and range-separated~\cite{RShybridDFs_Gerber} functionals that incorporate some fraction of nonlocal Hartree-Fock (HF) exchange.

A conceptually different approach to energy bands in crystals rests on \emph{ab initio} wave-function (WF) electronic-structure theory.~\cite{Helgaker}
Many-body WF's are explicitly computed with this approach and the relevant information on the role of various correlation effects can be further mapped onto a QP model.
Well-defined and controllable approximations initially developed and for a long time tested in
molecular quantum chemistry (QC) are employed, which ensures a reliable and transparent description of the correlation hole of the QP.
Band structure calculations based on QC methods were mainly focused on nonmetalic systems
\cite{Sun, Ayala, Forner, Stoll1, Shukla2, Birken2, Gr1, Gr2, Albrecht, Birken1, GF2, Shukla1, Stollhoff, Horsch1, Suhai, Hirata, Gruneis, BezuglyJP}.
Either second-order M\o ller-Plesset theory (MP2),\cite{Sun, Ayala, Suhai, Gruneis}
coupled-cluster (CC) techniques,\cite{Forner} or an effective local Hamiltonian approach (LHA)
\cite{Horsch1, Gr1, Birken1, Birken2, Albrecht, BezuglyJP} were used.

With WF-based approaches to solids, it is crucial to make use of the local character of the SR part of the correlation hole, as it is done for example in the LHA~\cite{Gr1, Albrecht, Birken1, Birken2, GF2, Stollhoff} or in local MP2 treatments \cite{Pisani, Pisani2008}. Sets of real-space, localized orbitals are thereby employed, being either derived by various localization procedures \cite{Foster-Boys, Edmiston, Pulaypaos} from the canonical Bloch states~\cite{Pisani, Zikovich, disent, Marzari, Forner, Karin} or obtained directly
in the crystal HF calculation by using a Wannier representation and
real-space iterative procedures.~\cite{Shukla2, Shukla1, Malrieu}

While the LHA, in particular, has been earlier shown to provide results in good agreement with
the experiment for ionic (or partly ionic) insulators such as MgO, BN, ZnS, and TiO$_2$,~\cite{Hozoi,AlexJCP,AlexPRB,interface_new} the goal of the present work is to analyze to which accuracy WF-based methods and the local Hamiltonian model can be employed to determine QP bands of covalent systems.  
We select for this purpose diamond, a prototype \emph{sp} insulator that has long served to
demonstrate the state of the art of first-principle formalisms.
Both valence- and conduction-band states are treated on equal footing in the present study,
in contrast to earlier QC investigations in the LHA framework that mainly focused on the valence-band physics.\cite{Albrecht, Gr1, Gr2}

The paper is organized as follows. In Section \ref{SecII_Quasi}, the description of QP's in the WF QC framework  is briefly recapitulated. Sections \ref{SecIII_Methods} and Appendix \ref{technicaldetails} contain information on the method, material model, and computational details. QP band-structure results for diamond are presented in Section \ref{Correlbands}, followed by conclusions in Section \ref{Concl}.

%%%%%%%%%%%%%%%END OF  INTRODUCTION%%%%%%%%%%%%%%%%%%%%%%

\section{Quasiparticle states}\label{SecII_Quasi}

\begin{figure}[!tb]
\caption{\label{PolRelcloudSDCI} Sketch of (a) the relaxation and polarization cloud generated around an extra charge placed into a ``frozen'' WO  $w_{n\sigma}$ with lattice vector $\mathbf{R}$ when a SCF approximation is made and (b) the same cloud plus quantum fluctuations, with the latter depicted in form of a Feynman diagram.}
\includegraphics[width=8.5cm]{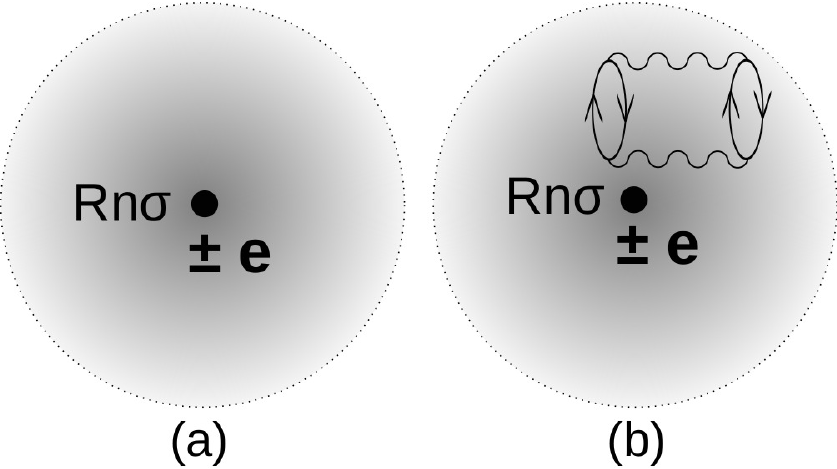}
\end{figure}

The starting point in our investigation is a HF calculation for the extended solid. ($N$--$1$)- and ($N$+$1$)-particle states are next constructed in a real-space representation in
terms of HF Wannier-type orbitals (WO's). A conceptually attractive way of describing the modifications of the surroundings when an extra particle, hole or electron, is added to the system is to carry out additional self-consistent-field (SCF) 
optimizations for the ($N\!\pm\!1\!$) states. In these SCF optimizations, we keep the WO hosting the extra particle ``frozen''. The outcome of such a SCF calculation is illustrative of relaxation and polarization effects in response 
to the extra charge, i.e., describes the static part of the correlation hole.
The basic idea is sketched in Fig.~\ref{PolRelcloudSDCI}.
%%%%%%%%%%%%%%%%
% FIGURE 1 %
%%%%%%%%%%%%%%%%
In practice, the additional SCF computation is carried out for a finite number of atoms and the 
technicalities are discussed in the next sections.

\begin{widetext}
\begin{figure}[!tb]
%\begin{center}
\caption{\label{LGSC}Diagramatic representation of the loss of ground-state correlation.}
\includegraphics[width=17.0cm]{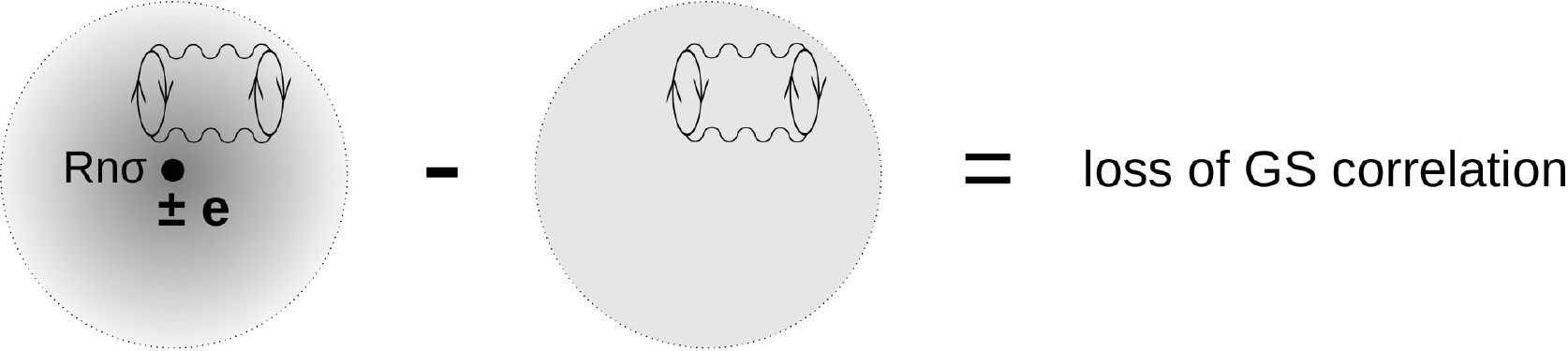} 
%\end{center}
\end{figure}
\end{widetext}

In addition to the static part of the correlation hole, quantum fluctuations may also give rise to sizable corrections. Quantum fluctuation effects account for the physical fact that, for example, the polarizability of a bond differs when electrons in the bond are treated at the mean-field level from the case when their correlations are taken into account. In the latter type of treatment, the polarizability is usually reduced. Quantum fluctuations are also present in the absence of the extra particle, i.e., in the GS of the $N$-particle system, where they contribute strongly, for example, to the binding energy. The presence of an extra electron or hole in a ``frozen" WO reduces the quantum fluctuations in its vicinity. This is due to blocking of some (one- and two-particle) excitations which are possible in the GS of the \emph{N}-particle system. Changes in the fluctuations when adding/removing an electron as compared to fluctuations in the GS are termed loss of ground-state correlations (LGSC). Quantum fluctuations in the presence of the extra particle are symbolically indicated by a Feynman diagram in Fig.~\ref{PolRelcloudSDCI}.
The LGSC is illustrated in Fig.~\ref{LGSC}. In practice, the fluctuations are computed by either configuration interaction (CI, be it not size consistent, see Sect.~\ref{SecIII_Methods})~\cite{Birken1, Birken2, Hozoi, AlexJCP, AlexPRB}, coupled cluster (CC, size-consistent)~\cite{Stoll_diamond}, or quasi-degenerate variational perturbation theory (size-consistent)~\cite{Cave, Gr1}. 

As compared to the HF data, relaxation and polarization effects within the crystalline surroundings of the added charge and additional correlation effects substantially reduce the energy required for adding or removing an electron.
Consequently, the band gap is also strongly reduced if such effects are properly accounted for.
Significant ``shrinking'' may further occur for the widths of the valence and conduction bands,
which translates into a larger QP effective mass. The band-width contraction and the LGSC lead to additional fine tuning of the band structure, i.e., an increase in the band gap's value. The most attractive feature of the WF QC approach is that each of these contributions can be separately analyzed. In the following sections, we aim to show to what extent accuracy can be achieved and where further progress is required.

%%%%%%%%%%%%%%%%
%%% FIGURE 2 %%%
%%%%%%%%%%%%%%%%
% 
%%%%%%%%%%%%%%%Section II: Method and model %%%%%%%%%%%%%%%
\section{Method and model}\label{SecIII_Methods}
\subsection{Basic relations}\label{QuasiP}
By adopting the QP picture of Sect.~\ref{SecII_Quasi}, the correlated $sp$ valence- and conduction-band energies are defined through 
\begin{eqnarray}\label{epsiloncorr}
\epsilon_{\mathbf{k}\mu\sigma}&=&E^{N}_{0}-\langle \Psi_{\mathbf{k}\mu\sigma}^{N-1}|H|\Psi_{\mathbf{k}\mu\sigma}^{N-1}\rangle,  
\nonumber \\
\epsilon_{\mathbf{k}\nu\sigma}&=&\langle \Psi_{\mathbf{k}\nu\sigma}^{N+1}|H|\Psi_{\mathbf{k}\nu\sigma}^{N+1}\rangle-E^{N}_{0},
\end{eqnarray}
where $E^{N}_{0}$ is the energy of the correlated GS of the $N$-electron crystal.
The many-body ($N\!+\!1\!$)- and ($N\!-\!1\!$) states $|\Psi_{\mathbf{k}\nu\sigma}^{N+1}\rangle$ and $|\Psi_{\mathbf{k}\mu\sigma}^{N-1}\rangle$, respectively, describe the
QP states with spin $\sigma$  
and crystal momentum $\mathbf{k}$. These ($N\!+\!1\!$) and ($N\!-\!1\!$) states are associated with conduction- and valence-band indices $\nu$ and $\mu$. The Hamiltonian
matrix elements (ME's) in Eq.~(\ref{epsiloncorr}) refer to the QP energies (see below). 
We focus here on the ($N\!+\!1\!$) states, the case of the ($N\!-\!1\!$) states being analogous. The ($N\!+\!1\!$) states $|\Psi_{\mathbf{k}\nu\sigma}^{N+1}\rangle$ can be described by the ansatz \cite{Fuldebook, Fuldebook2}
\begin{eqnarray}\label{Quasistate}
|\Psi_{\mathbf{k}\nu\sigma}^{N+1}\rangle&=&\Omega c^{\dagger}_{\mathbf{k}\nu\sigma} |\Phi_{\textsc{SCF}}\rangle\,, \nonumber\\
\Omega&=&e^{S}, 
\end{eqnarray}
where $|\Phi_{\textsc{SCF}}\rangle$ denotes the $N$-particle
HF GS. 
The operator $c^{\dagger}_{\mathbf{k}\nu\sigma}$ creates an extra electron in a HF Bloch state with index $\nu$, momentum $\mathbf{k}$, and spin $\sigma$. 
The HF ($N\!+\!1\!$) state
$\vert\Phi_{\mathbf{k}\nu\sigma}^{N+1}\rangle$=$c^{\dagger}_{\mathbf{k}\nu\sigma} |\Phi_{\textsc{SCF}}\rangle$
in Eq.~(\ref{Quasistate}) is transformed by the wave (or M\"oller) operator $\Omega$ into the 
correlated $\vert\Psi_{\mathbf{k}\nu\sigma}^{N+1}\rangle$ state. Here, we assume a QP approximation, which implies that for each SCF conduction-band state $\vert\Phi_{\mathbf{k}\nu\sigma}^{N+1}\rangle$ there exists a corresponding correlated state  $\vert\Psi_{\mathbf{k}\nu\sigma}^{N+1}\rangle$ in the interacting ($N\!+\!1\!$) system. For $\Omega$, we for the moment choose
a CC-like ansatz, with $S$ being generally a scattering operator, see, e.g., Refs.~\onlinecite{Fuldebook, Fuldebook2}. The transformation between Bloch and Wannier states is 
\begin{eqnarray}\label{BlochWannier}
c^{\dagger}_{\mathbf{k}\nu\sigma}=\frac{1}{\sqrt{N_{B}}} \sum_{n, \mathbf{R}_{I}}\alpha_{\nu n}(\mathbf{k})e^{i\mathbf{k}.\mathbf{R}_I}
w^{\dagger}_{\mathbf{R}_In\sigma}\,,
\end{eqnarray}
where the local operator $w^{\dagger}_{\mathbf{R}_In\sigma}$ creates an electron in the WO $|w_{n\sigma} (\mathbf{R}_{I})\rangle$ of index $n$ and spin $\sigma$,
centered at a site with lattice vector $\mathbf{R}_I$ of the unit cell \emph{I}. $N_{B}$ is the number of unit cells.
The local ($N\!+\!1\!$) SCF state can be then described in
direct space as
 \begin{eqnarray}\label{phinplus1}
|\Phi^{N+1}_{\mathbf{R}_{I}n\sigma}\rangle=w^{\dagger}_{\mathbf{R}_{I}n\sigma} |\Phi_{\textsc{scf}}\rangle. 
\end{eqnarray}
The $\alpha_{\nu n}(\mathbf{k})$ matrix is determined by the crystal structure, here the diamond lattice. In fact, the choice of $\alpha_{\nu n}(\mathbf{k})$ depends on the localization scheme for the Wannier functions. In the present study, the Wannier-Boys algorithm~\cite{Zikovich} is employed for the optimization of the unitary band-mixing matrix $\alpha_{\nu n}(\mathbf{k})$ in the multi-band Wannier transformation. 

By inserting Eqs.~(\ref{Quasistate}) and~(\ref{BlochWannier}) in Eq.~(\ref{epsiloncorr}), the QP conduction bands are expressed as 
\begin{eqnarray}
\epsilon_{\mathbf{k}\nu\sigma}&=&
\sum_{\mathbf{R}_{I}}\sum_{nn'}\alpha_{\nu n}(\mathbf{k})\alpha^{*}_{\nu n'}(\mathbf{k})e^{i\mathbf{k}\,.\mathbf{R}_{I}} 
\nonumber \\
&\times& 
\langle\Phi_{\textsc{scf}}|w_{\mathbf{0}n'\sigma}e^{S^{\dagger}}H e^{S}w^{\dagger}_{\mathbf{R}_{I}n\sigma}|\Phi_{\textsc{scf}}\rangle -E^{N}_{0}. 
\label{epsiloncorr2}
\end{eqnarray}
Here $E^{N}_{0}=\langle\Phi_{\textsc{scf}}|e^{S_{0}^{\dagger}}He^{S_{0}}|\Phi_{\textsc{scf}}\rangle$ and the wave operator $\Omega_{0}$=$e^{S_{0}}$ transforms $|\Phi_{\textsc{scf}}\rangle$ into the ``true'' GS $|\Psi^{N}_{0}\rangle$ of the $N$-particle system. Equation (\ref{epsiloncorr2}) contains the assumption that the matrix $\alpha_{\nu n}(\mathbf{k})$ remains unchanged by correlation effects, which is a reasonable assumption for quasiparticles. 
It is easily seen from that equation that the energies $\epsilon_{\mathbf{k}\nu\sigma}$ are expressed in terms of ME's of a local effective Hamiltonian $H^{\rm eff}$ operator
\begin{eqnarray}\label{Heff}
H^{\rm eff}=PHP-E^{N}_{0}P\,, \nonumber \\
P=\sum_{\mathbf{R}_K n' \sigma} |\Psi_{\mathbf{R}_Kn'\sigma}^{N+1}  \rangle \langle\Psi_{\mathbf{R}_Kn'\sigma}^{N+1}| 
\end{eqnarray}
defined within a set of direct-space, ``locally correlated"
($N\!+\!1\!$) states
\begin{eqnarray}\label{DScorrelWF}
|\Psi^{N+1}_{\mathbf{R}_{K}n'\sigma}\rangle=e^{S}w^{\dagger}_{\mathbf{R}_{K}n'\sigma}|\Phi_{\textsc{scf}}\rangle\,.
\end{eqnarray}
For relatively weakly correlated systems like diamond, where the amplitude of the bare particle in the QP is not too small, one may limit the operator \emph{S} to include
only one- and two-particle excitations out of the space of occupied spin-orbitals~\cite{Fulde, Fuldebook, Birken2, Gr1, Borrmann}. These excitations are further confined to excitations out of the space of occupied localized SCF orbitals, when a QC program package such as \textsc{molpro} \cite{molpro} is used for the computation of the correlation hole. 

The correlation corrections to the conduction bands follow from
\begin{displaymath}
\epsilon^{\rm corr}_{\mathbf{k}\nu\sigma}=\epsilon_{\mathbf{k}\nu\sigma}-\epsilon^{\textsc{scf}}_{\mathbf{k}\nu\sigma}\,.
\end{displaymath}
In analogy to Eq.~(\ref{epsiloncorr2}), we rewrite this energy difference in terms of local ME's as
\begin{eqnarray}
\epsilon^{\rm corr}_{\mathbf{k}\nu\sigma}&=&
\sum_{\mathbf{R}_{I}}\sum_{nn'}\alpha_{\nu n}(\mathbf{k})\alpha^{*}_{\nu n'}(\mathbf{k})e^{i\mathbf{k}.\mathbf{R}_{I}} 
\nonumber \\
&\times& 
\langle\Phi_{\textsc{scf}}|w_{\mathbf{0}n'\sigma}(e^{S^{\dagger}}H e^{S}-H)w^{\dagger}_{\mathbf{R}_{I}n\sigma}|\Phi_{\textsc{scf}}\rangle \nonumber \\
&-&(E^{N}_{0}-E_{0}^{\textsc{scf}}),
\label{epsiloncorr3}
\end{eqnarray}
where the GS energy $E_{0}^{\textsc{scf}}$ of the $N$-electron system 
system as well as the WO's are obtained from a periodic HF SCF calculation. 
The expression in Eq.~(\ref{epsiloncorr3}) can be rewritten in an elegant form by using
cumulants and we refer to Refs.\,\onlinecite{Fulde,Fuldebook,Fuldebook2} for further details.
The main point to note is that for the computation of the local ME's in Eq.~(\ref{epsiloncorr2})
it is not necessary to consider an infinite system, due
to the predominantly local character of the SR part of the correlation hole. 
All considerations can be applied instead to properly embedded finite clusters whose size depends on the physics and accuracy one aims to achieve (see Sect.~\ref{SecIIIB_clustersandwavefunctions}).  
Obviously, similar relations as those in equations (1--7) hold for the valence-band ($N\!-\!1\!$) states and energies. 
\subsection{Computational strategy}\label{SecIIIB_clustersandwavefunctions}
The starting point in the calculation of the correlated valence- and conduction-band states is the HF GS $|\Phi_{\textsc{scf}}\rangle$. The Bloch orbitals obtained from the periodic SCF calculation are in a first step submitted to a Wannier-Boys localization procedure \cite{Zikovich} that yields valence- and conduction-band WO's. The Wannier-Boys localization is applied separately to the HF occupied (i.e., core and valence) and lower-lying, unoccupied (i.e., lower-lying conduction-band) Bloch states of the neutral crystal. In diamond, the optimally localized valence- and conduction-band WO's are two-center C--C bonding and antibonding orbitals. By adding an electron to the (antibonding) WO centered at bond $n$ in unit cell $I$, the part $\vert\Phi^{N+1}_{\mathbf{R}_{I}n\sigma}\rangle$=$w^{\dagger}_{\mathbf{R}_{I}n\sigma}|\Phi_{\textsc{scf}}\rangle$ in Eq.~(\ref{DScorrelWF}) can be considered as known. In a next step, the prefactor $e^{S}$ (and $|\Psi^{N+1}_{\mathbf{R}_{I}n\sigma}\rangle$) is determined, that describes the correlation hole around the added electron and  is needed in order to compute the ME's in Eq.~(\ref{epsiloncorr2}). It suffices to compute thereby only symmetry non-equivalent ME's. Given the local character of the SR part of the correlation hole, it is further sufficient to consider an adequately large and properly embedded cluster, cut off from the extended solid. This SR part of the correlation hole is thus explicitly dealt with in post-HF, embedded-cluster QC calculations. 
The LR part of the polarization cloud is evaluated separately within the dielectric continuum approximation (see Sect.~\ref{diag_correlH}). The same holds for the hole states $\vert\Psi^{N-1}_{\mathbf{R}_{I}m\sigma}\rangle$.

\begin{figure}[!tb]
\begin{center}
\caption{\label{Model_C} Sketch for the partition of a given embedded cluster $\mathcal{C}$ into the active $\mathcal{C_A}$ 
and buffer $\mathcal{C_B}$ regions.
}
\includegraphics[width=8.5cm]{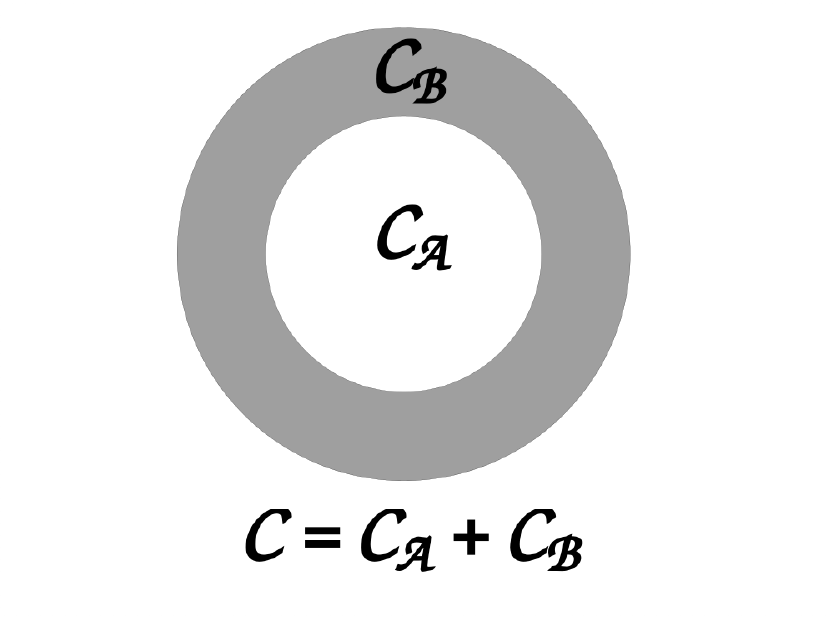}
\end{center}
\end{figure}

In the following, we consider a bond $n$ situated in the center of a given cluster $\mathcal{C}$ and defined by
the lattice vector $\mathbf{R}_{I}$.
For the post-HF calculations, the cluster $\mathcal{C}$ is divided into a central ``active'' subunit $\mathcal{C_A}$, for which the actual correlation treatment is carried out, and a buffer region $\mathcal{C_B}$ (see Fig.~\ref{Model_C}).
%%%%%%%%%%%%%FIGURE 3%%%%%%%%
The atomic sites in the buffer $\mathcal{C_B}$ region and the associated Gaussian-type-orbital (GTO) basis functions are needed for properly describing the tails of WO's centered within the $\mathcal{C_A}$ domain. Such tails of Wannier functions localized on ``active'' C--C bonds have to be adequately treated when relaxation and polarization in $\mathcal{C_A}$ are computed (see Appendix~\ref{AppA}). The tails of WO's centered in $\mathcal{C_B}$ that reach outside the cluster $\mathcal{C}$ need also special care. Cutting off these tails is rigorously done by means of a projection of the crystal WO's onto the set of GTO basis functions within $\mathcal{C}$, see Appendix~\ref{AppA} and Refs.~\onlinecite{Birken2, Hozoi, AlexJCP, AlexPRB}. Projected WO's are hence employed in the correlation treatment. 
All WO's centered in $\mathcal{C_B}$ remain inactive in the correlation calculation. Therefore, their less accurate representation is not of primary importance. 

The crystal environment $\mathcal{E}$ is explicitly included in the cluster correlation calculations as an effective one-electron embedding potential $V^{\rm emb}$.
$V^{\rm emb}$ incorporates the effect of the frozen occupied HF orbitals outside $\mathcal{C}$ on the electrons within $\mathcal{C}$ as well as the interaction of the nuclei of $\mathcal{E}$ with electrons in $\mathcal{C}$.
Since the WO's within $\mathcal{C_B}$ are kept frozen all the time, the Coulomb and exchange potentials associated with those are also incorporated in $V^{\rm emb}$.

When determining the prefactor $e^{S}$ in Eq.~(\ref{DScorrelWF}), we consider excitations out of bonding WO's centered in the $\mathcal{C_A}$ region.
The virtual space into which electrons are excited consists of projected atomic orbitals (PAO's). The PAO's are generated from the GTO basis functions centered within $\mathcal{C_A}$ after projecting out\cite{Hampel} the core, valence-band, and lower-lying conduction-band WO's of $\mathcal{C}$ (see Appendix~\ref{AppA}). The PAO's are orthogonal to those WO's by construction. 

We note that the local creation and annihilation operators in Eq.~(\ref{epsiloncorr2}) refer to spin orbitals. In the actual QC calculations, however, spatial orbitals are used.
While in a first set of SCF optimizations for the $(N\!\pm\!1)$ states we ``freeze'' the orbital hosting the extra particle, hole or electron, and account for relaxation and polarization of all other occupied orbitals within the $\mathcal{C_A}$ region, both on-site and at a number of nearby neighbors around the ``frozen'' orbital, in a second SCF run we allow for relaxation of the singly occupied orbital and keep the static part of the correlation hole (the previously relaxed orbitals) fixed~\footnote{The bonding WO's hosting an extra hole do not delocalize in those SCF orbital optimizations. On the other hand, keeping the antibonding $(N\!+\!1)$ WO's localized in that kind of additional SCF optimizations is not possible. The corresponding relaxation effect is, however, expected to be of the same magnitude as for the ($N\!-\!1\!$) states}.
One can view this as accounting for spin dependence of the correlation hole. 

Quantum fluctuations are finally described by CI calculations with single and double excitations (CISD)~\cite{ICCI}.
The two-particle excitations modify somewhat the charge distribution since the polarizability of a bond differs when the electrons in this bond are treated as being either independent or correlated. This readjustment of the density distribution can be also described by single-particle excitations, see Appendix~\ref{AppLGSC}.
%%%%%%%%%%%%End of Section II: Method and models%%%%%%%%%%%%
%
%%%%%%%%%%%%%%%%%Section results%%%%%%%%%%%%%%%%%%

\section{Hartree-Fock band structure}\label{HF_bands}
\subsection{{\sc crystal} HF bands}
The HF band structure of diamond, computed using the {\sc crystal} program \cite{CRYSTAL1}
and triple-zeta GTO basis sets \cite{Dunning, Birken2} (see Appendix \ref{technicaldetails}), is shown in Fig.~\ref{hfband_new} with solid lines.
%
%%%%%FIGURE 4%%%%%%%%%%%%%%%
\begin{figure}[!tb]
\begin{center}
\caption{\label{hfband_new} HF band structure of diamond along different high-symmetry directions from both periodic (\textsc{crystal}) calculations (solid lines) and embedded cluster calculations (dashed lines).
}
\includegraphics[width=8.5cm]{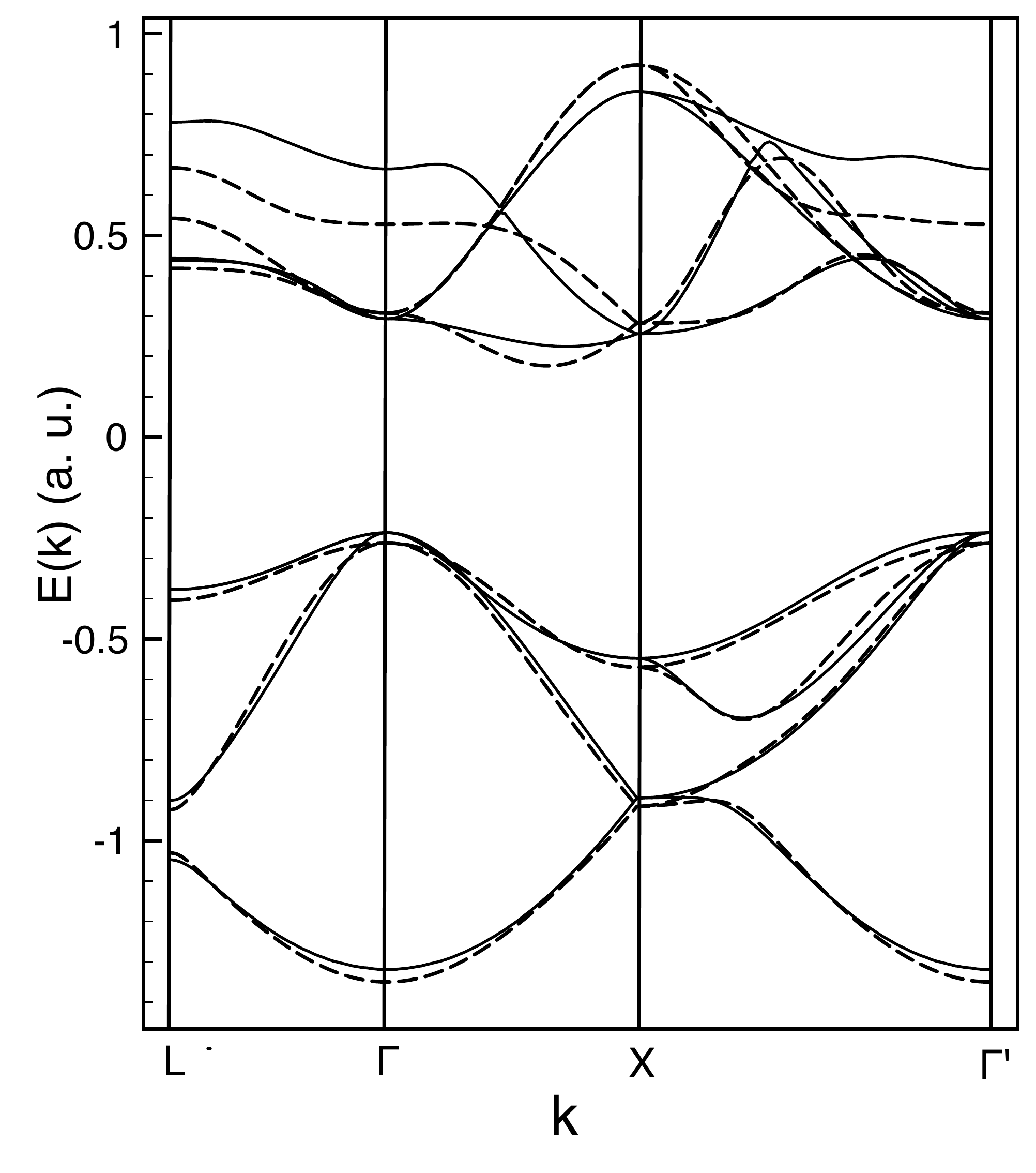}
\end{center}
\end{figure}
The HF value for the width of the hybridized $sp$ valence bands, 29.4 eV, considerably overestimates the experimental value of 23.0$\pm$0.2 eV~\cite{experBWdiamond}. 
The fundamental gap is indirect, with the top of the valence bands at the $\Gamma^{v}_{25^{'}}$ point
and the bottom of the conduction bands along the $\Gamma$$\rightarrow$X ($\Delta$) symmetry line.
In our HF calculation it comes out as 12.6 eV, much larger than the experimentally derived value of 5.5 eV~\cite{Clark}.
The HF direct band gap ($\Gamma^{v}_{25^{'}}$$\rightarrow$$\Gamma^{c}_{15}$) is 14.4 eV, as compared to an experimental estimate of 7.3 eV~\cite{Philipp}.   
Localized WO's are separately obtained for the core, valence, and lowest-lying conduction bands. Since the first four low-lying conduction bands are separated from the higher virtual Bloch states by a finite gap, the Wannier-Boys transformation is easily carried out for those. It is not the case for silicon, for example, where the lower-lying conduction bands are entangled with higher-lying virtual states and special localization techniques need to be applied~\cite{disent}. 

\begin{figure}[!tb]
\begin{center}
\caption{\label{ProjWOs}$sp^{3}$ type WO's for the conduction ({\it a}) and valence ({\it b}) bands of diamond
after projection onto a [C$_{56}$] cluster.}
\subfigure []{
\includegraphics[width=7.5cm]{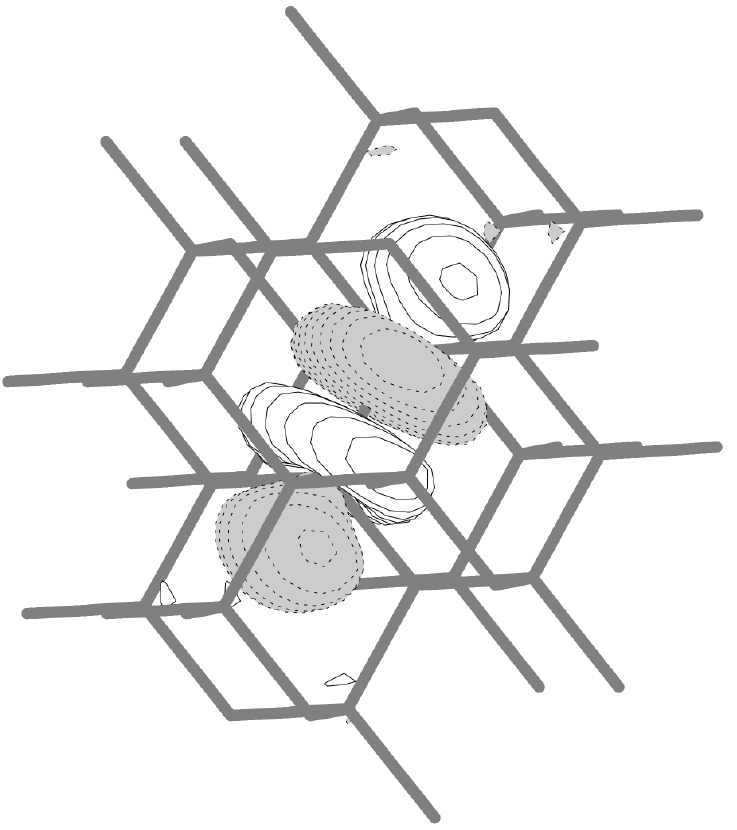} }
\subfigure[]{
\includegraphics[width=7.5cm]{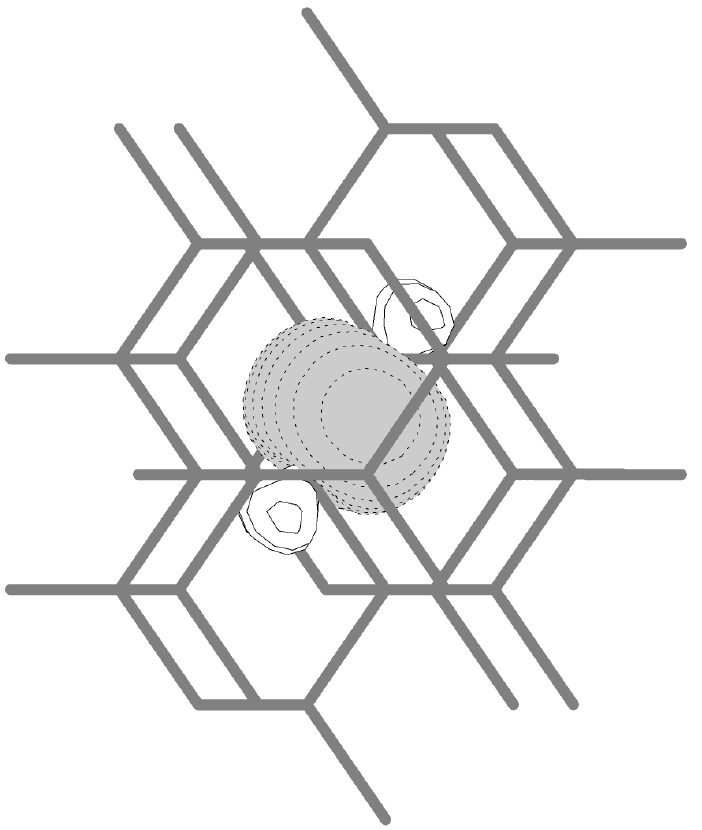}}
\end{center}
\end{figure}

The resulting valence-band and conduction-band WO's constitute a set of four bonding and four antibonding orbitals~\footnote{The bonding and antibonding WO's are constructed from $sp^{3}$-type
hybrid functions at the C sites. Each WO is oriented along one of the four C--C bonds in a C$_4$ tetrahedron.}.
Whereas the bonding WO's are centered and rather well localized at a given C--C bond, the antibonding WO's have non-negligible LR tails. This is illustrated in Fig.~\ref{ProjWOs}.
We analyzed the degree of localization of the WO's in terms of the atomic delocalization index defined in Ref.~\onlinecite{Zikovich}. The latter gives an estimate of the mean number of atoms that ``contribute" to a WO \cite{Zikovich, PM}. The four antibonding type WO's are characterized by atomic delocalization indices of 2.63 e$^{-2}$. This indicates that the WO's are each mainly centered on the two carbons forming a bond but additionally have substantial weight at NN and second NN C sites. 
%
%FIGURE 5 %%%%%%%%%%%%%%%%%%%
The atomic delocalization indices for the four bonding type WO's are 2.10 e$^{-2}$, which shows a faster decay away from the C--C bonds on which they are mostly localized. Mulliken population data \cite{Zikovich}, that also allow an estimate of the WO's spatial extent, are consistent with these findings (see Appendix \ref{AppA}). 
After projecting the bonding and antibonding WO's onto the set of basis functions of the finite embedded cluster $\mathcal{C}$, the norms of the projected bonding WO's within the $\mathcal{C_A}$ region are always larger than 0.99 of the original WO's. The norms of the projected antibonding WO's are in the range 0.92--0.99, never smaller than 0.92 even for the smallest clusters considered in the post-HF calculations.
\subsection{HF bands with projected WO's} \label{hffromcluster}

Given the close resemblance of the projected WO's to the initial HF WO's and the use of an accurate HF cluster-in-solid embedding scheme, the energy bands computed in terms of ME's of H$^{\rm eff}$ over the frozen-orbital (FO) restricted open-shell HF (ROHF) ($N\!\pm\!1\!$) states $\vert\Phi^{N\pm1}\rangle$ (Eqs.~(\ref{phinplus1}) and~(\ref{scfMEs})) will closely reproduce the bands from
the periodic HF calculation. For the ($N\!+\!1\!$) states, for example, the off-diagonal ME's are (see also Sect.~\ref{nondiagH})
\begin{eqnarray}\label{scfMEs}
H^{\textsc{scf}}_{\mathbf{R}_I, nn'}&=&
\langle\Phi_{\textsc{scf}}|w_{\mathbf{0}n'\sigma}Hw^{\dagger}_{\mathbf{R}_{I}n\sigma}|\Phi_{\textsc{scf}}\rangle-\delta_{\mathbf{R}_I\mathbf{0}}\delta_{nn'}E^{\textsc{scf}}_{0} \nonumber \\
\vert\Phi^{N+1}_{\mathbf{R}_In\sigma}\rangle&=&w^{\dagger}_{\mathbf{R}_{I}n\sigma}|\Phi_{\textsc{scf}}\rangle\,.
\end{eqnarray}
The bonds $I$ are here NN, 2nd NN, or 3rd NN bonds around the 0 ``reference'' bond. Their positions are defined by the lattice vectors $\mathbf{R}_I$. In a tight-binding picture, these HF ME's are essentially the \emph{bare} hopping integrals $t$ (see Eq.~(\ref{nonort_TB}))~\footnote{The HF off-diagonal ME's are obtained by means of FOCI calculations for the mutually orthogonal FO-ROHF
WF's $\Phi^{N\pm1}$}. 
We used a cluster of 184 carbon atoms for these FO-ROHF calculations for the $\vert\Phi^{N\pm1}\rangle$ states.
Diagonal ME's $H^{\textsc{scf}}_{\mathbf{0}, nn}$ are listed for both valence and conduction bands on the second lines of Tables~\ref{diagonalhole} and \ref{diagonalelectron}, respectively.
Off-diagonal HF ME's (see Eqs.~(\ref{scfMEs}) and ~(\ref{nonort_TB})) are reported in Table~\ref{Hop_bare_integrals}, up to 3rd neighbor bonds (ME's for which $I$ is between 3 and 6).  
\begin {table} [htbp]
\centering
\caption{Off-diagonal HF ME's, i.e, bare hopping integrals, for valence-band (t$_{mm'}$) and
conduction-band (t$_{nn'}$) states of diamond (in eV). There is more than one type of 2nd and 3rd neighbor bonds around the 0 ``reference'' bond.}
\begin{tabular}{p{1.3cm}p{0.8cm}p{1.3cm}p{1.4cm}p{1.3cm}p{1.0cm}p{1.0cm}p{1.0cm}p{1.0cm}} \hline\hline
&&NN&\multicolumn{2}{c}{2nd NN}&\multicolumn{4}{c}{3rd NN}\\ 
I&&1&2 trans&2 cis&3&4&5 &6\\ \hline
t$_{mm'}(\mathbf{R}_{I})$&&2.701& 0.882&0.486&0.230&0.074&0.173&0.046 \\ \\
t$_{nn'}(\mathbf{R}_{I})$&&0.242&0.692&0.408&0.939&0.433&0.125& 0.310\\
 \hline\hline
\end{tabular} 
\label{Hop_bare_integrals}
\end {table}

Having all these data, the real-space matrix of the effective Hamiltonian can be recast in $\mathbf{k}$-space and the corresponding matrix $H(\mathbf{k})$ can then be easily diagonalized. The resulting band structure is shown with dashed lines in Fig.~\ref{hfband_new}. As also found in, e.g., Refs.~\onlinecite{Gr1, Gr2, Albrecht}, including off-diagonal ME's up to 3rd NN's is sufficient to reasonably well reproduce the valence bands of the periodic calculation. 
This is not surprising, given the fast decay of the valence-band off-diagonal ME's, see second line in Table~\ref{Hop_bare_integrals}. Yet, small deviations from the \textsc{crystal} HF bands are also observed in some particular regions in $k$-space. For example, the \textsc{crystal} band structure indicates
smaller effective masses at the $\Gamma^{v}_{25'}$ point, just as for Si \cite{Gr1}. Shifts to lower energies by 0.7--0.9 eV are further observed at the top ($\Gamma^{v}_{25'}$) and bottom ($\Gamma^{v}_{1}$) of the valence bands derived from the embedded cluster calculations. 
The valence band width $\Delta_{\rm {VBW}}$ is thus to some extent affected -- it now amounts to 29.6 eV, as compared to a value of 29.4 eV in the initial periodic HF calculation.

Clearly, the neglect of off-diagonal ME's beyond the 3rd NN bonds is somewhat more significant for the conduction-band states (see Fig.~\ref{hfband_new}), due to the larger spatial extent of the antibonding WO's.
The values listed for the conduction-band hoppings in Table~\ref{Hop_bare_integrals} show that
the NN, 2nd neighbor, and some of the 3rd neighbor ME's are in fact of similar magnitude. 
The bottom of the conduction bands $\Delta_{\rm min}$, for example,
is now lower by about 1.3 eV as compared to the \textsc{crystal} $\Delta_{\rm min}$ and is shifted closer to the $\Gamma$ point along the $\Delta$ ($\Gamma$\,$\rightarrow$\,X) symmetry line. 
As a result, the indirect band gap $\Gamma^{v}_{25'}$--$\Delta^{c}_1$ is 11.9 eV, smaller than
the \textsc{crystal} value of 12.6 eV.
The direct band gap $\Gamma^{v}_{25'}$ --$\Gamma^{c}_{15}$ is on the other hand somewhat enlarged to 15.5 eV,
due to an upward shift of the conduction bands by 0.4 eV at the $\Gamma^{c}_{15}$ point.
The deviations between the {\sc crystal} conduction bands and the conduction bands obtained from the cluster calculations are somewhat more pronounced along the $\Gamma$$\rightarrow$L (close to the L point) and $\Gamma$$\rightarrow$X symmetry lines. For example, the doubly-degenerate $sp$ conduction bands along  $\Lambda$ ($\Gamma$$\rightarrow$L), whose wave functions transform according to the irreducible representation $\Lambda_3$ as well as the $sp$ band $\Lambda_{2}$ (L$_{2'}$ at L) show somewhat different energy dispersions in the {\sc crystal} and cluster calculations (solid versus dashed lines in Fig.~\ref{hfband_new}).

Interestingly, we found that the NN and 2nd NN off-diagonal HF ME's for the valence-band
states in Table~\ref{Hop_bare_integrals} differ by only $\approx$0.01 eV from those obtained by Gr\"afenstein \emph{et al.} \cite{Gr1, Gr2} from calculations on hydrocarbon molecules (see Table I in Ref.~\onlinecite{Gr1} and Table IV in  Ref.~\onlinecite{Gr2}).
Slightly larger differences, in the range of 0.05 eV, are observed for the 3rd neighbor ME's. 
On the  other hand, an accurate description of the conduction electrons requires  a careful 
treatment of the periodic potential. This is here achieved with our recently designed cluster-in-solid HF embedding scheme.
Hydrogen-terminated carbon clusters as used in several earlier studies~\cite{Gr1, Gr2, Albrecht}
provide electron affinities very much different as compared to the crystalline material
and the extra electron would be at best only loosely bound to such a cluster. 
\section{Quasiparticle bands}\label{Correlbands}
\subsection{Correlation corrections to the diagonal matrix elements} \label{diag_correlH}
\begin{figure}[!tb]
\begin{center}
\caption{\label{clusterC110} Sketch of the [C$_{110}$] cluster used in the calculations.
The $\mathcal{C_{A}}$ region consists of a [C$_{2}$C$_{6}$C$_{18}$] fragment.
The two central active atoms defining the reference $(N\!\pm\!1)$ C--C bond are shown as large black spheres,
the six NN's as hatched white spheres, and the other 18 C atoms in $\mathcal{C_{A}}$ as white
``equatorial'' spheres. Sites in $\mathcal{C_{B}}$ are drawn as average-size black spheres.}
\includegraphics[width=8.5cm]{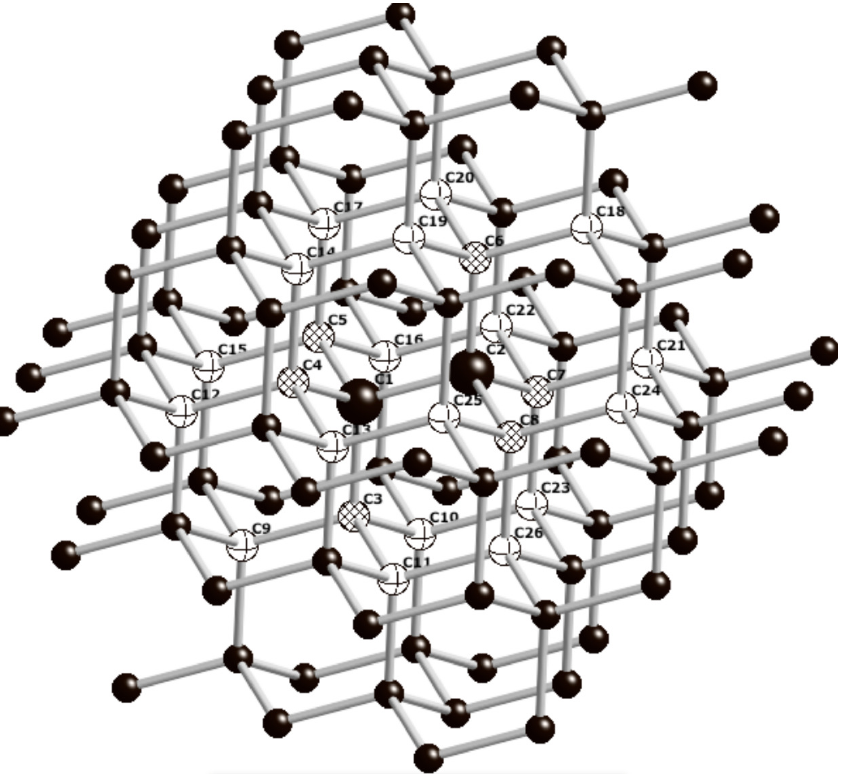}
\end{center}
\end{figure}

Correlation-induced corrections to the diagonal ME's are discussed in this section for both valence- and conduction-band states. To analyze in more detail the contribution of individual ``shells'' to the overall relaxation and polarization effects, we carry out calculations on a few different clusters. In addition to the [C$_{184}$] fragment employed in Sect.~\ref{hffromcluster} for extracting the HF ME's, embedded clusters containing either
56, 110, or 294 C sites are used.
The corresponding active regions, explicitly dealt with in the correlation calculations, comprise for the latter 8, 26, or 110 atoms, respectively. The active region of the [C$_{184}$] cluster contains 56 C sites.  
We label these clusters as [C$_{56}$], [C$_{110}$] (see Fig.~\ref{clusterC110}), and [C$_{294}$]. 

\paragraph {Static part of the correlation hole and quantum fluctuations.} 
%%%%%%%%%%Figure 6%%%%%%%%%%%
We first consider the static part of the correlation hole, with contributions up to the fourth
shell of neighboring C--C bonds around the reference ``frozen" WO hosting the extra hole ($(N\!-\!1\!)$ states) or electron ($(N\!+\!1\!)$) states). The static part of the correlation hole can be obtained by reducing the operator $S$ in Eq.~(\ref{DScorrelWF}) to only single-particle excitations around the ``frozen'' WO. These excitations generate dipole moments within the nearby bonds (modify the densities in the surroundings), describing thus relaxation and polarization effects. These latter effects are however assessed by means of additional ROHF calculations in our present study~\cite{Birken2, Hozoi, AlexJCP}. For each of the clusters we use, all doubly-occupied orbitals within the $\mathcal{C_{A}}$ region are here allowed to relax and polarize. 
As discussed in Ref.~\onlinecite{Pahl}, to first-order perturbation theory, the resulting ROHF WF's
$\vert\tilde{\Phi}^{N\!\pm\!1}\rangle$ for the ($N\!\pm\!1$) states are identical to the WF's
$\vert\tilde{\Psi}^{N\!\pm\!1}\rangle$ correlated only through single-particle excitations around the frozen $(N\!\pm\!1)$ WO.
Following Pahl \emph{et al.}~\cite{Pahl}, the approximation of keeping the extra particle and its orbital ``frozen" is termed  
as the frozen local hole approximation (FLHA). 

Various contributions from the nearby surroundings to the static part of the correlation hole are listed for the valence $(N\!-\!1)$ and low-lying conduction-band $(N\!+\!1)$ states in Table~\ref{diagonalhole} and Table~\ref{diagonalelectron}, respectively. The effect of quantum fluctuations, i.e., LGSC's, is also shown in those tables, as computed on a [C$_{56}$] cluster.
As expected, the relaxation and polarization effects at the shell of six NN C--C bonds are 
the largest, about --2.3 eV for electron-removal (Table~\ref{diagonalhole}, top) and about --1.0 eV for electron-addition (Table~\ref{diagonalelectron}, top) states.
The relaxation of the reference valence-band WO hosting the hole brings an additional energy gain of  --0.25 eV (see Table~\ref{diagonalhole}). This is related to the response of the electron with opposite spin that is accommodated into the same spatial orbital.

When an extra electron is added to the frozen antibonding orbital, the energy gain due to
relaxation and polarization of the corresponding doubly-occupied bonding WO (intrabond charge rearrangement), centered at the same C--C unit, amounts to about --0.28 eV (see Table~\ref{diagonalelectron}).
The sum contribution of the NN shell and intrabond charge rearrangement amounts
then to approximately --1.3 eV for the conduction-band states. This result agrees well with preliminary calculations by Birkenheuer \emph{et al.} \cite{Birken2} using the same cluster-in-solid embedding technique and LHA but a smaller [C$_{26}$] cluster with an active [C$_{8}$] region. The polarization of the 2nd and 3rd NN bonds was treated however within the dielectric continuum approximation in that earlier work, a rather crude approximation at close distances from the additional localized charge.
\begin{table}[htbp]
 \centering 
\caption{
Correlation-induced corrections $\Delta$H$_{mm}$($\textbf{0}$) to the diagonal Hamiltonian ME's for localized
valence-band hole states (in eV).
Negative corrections indicate an upward shift of the valence bands.
}
\begin{tabular}{p{4.3cm}p{1.4cm}p{1.4cm}p{1.4cm}p{1.4cm}p{0.01cm}}
  \hline\hline
Cluster &[C$_{56}$] &[C$_{110}$]&[C$_{184}$]&[C$_{294}$] \\ \hline 
HF &--18.78&--18.78&&\\ 
&&$\Delta$H$_{mm}$($\textbf{0}$)&& \\ \hline 
\emph{Static correlation hole} &&&&&\\
NN shell relaxation&--2.30 &--2.31&--2.31&--2.31\\
2nd NN shell  &&--0.98&--0.97&--0.97\\
3rd NN shell  &&&--0.62&--0.60\\
4th NN shell  &&&&--0.38\\
Reference-WO relaxation&--0.27&--0.25&--0.25&--0.25 \\
\emph{Long-range polarization} &--2.65&--1.81&--1.35&--1.08\\
\emph{Quantum fluctuations} &+0.91&+0.91&+0.91&+0.91 \\
Total correction&--4.31&--4.44&--4.59&\textbf{--4.68}\\ \hline \hline
 \end{tabular} 
\label{diagonalhole}
\end{table}

Charge relaxation and polarization effects at the 2nd NN bonds decrease by a factor of 1.5 to 2 as compared to the NN bond contribution, for both hole and electron-addition states.
A similar trend is observed for the relaxation and polarization of the 3rd and 4th shells of C--C bonds as compared to the contributions of the 2nd and 3rd shells, respectively.  
In contrast to partially ionic compounds such as BN~\cite{AlexJCP} and ZnS~\cite{AlexPRB},
relaxation and polarization of the 3rd and 4th coordination shells are in diamond larger.

The ROHF corrections to the diagonal ME's due to the relaxation and polarization effects
are overestimated because of the neglect of quantum fluctuations.
Concerning the latter, only those need to be considered which we termed above loss of ground-state correlations. 
Details on the explicit computation of the LGSC's are given in Appendix~\ref{AppLGSC}. These correlations are assessed by means of CISD calculations for the $N$-particle GS and for the ($N\!-\!1\!$) and ($N\!+\!1\!$) states. The CISD treatment is carried out on the [C$_{56}$] cluster, using the ROHF WF's as reference, and within the FLHA.
Results for the effect of quantum fluctuations, i.e., LGSC, as obtained from the differences
between the GS and $(N\!-\!1)$/$(N\!+\!1)$ CISD correlation energies, are listed in Tables~\ref{diagonalhole} and~\ref{diagonalelectron} and amount to 0.91 and 0.30 eV for the ($N\!-\!1\!$) and ($N\!+\!1\!$) states, respectively.
 \begin {table}[htbp]
\centering 
\caption{Correlation-induced corrections $\Delta$H$_{nn}$($\textbf{0}$) 
to the diagonal Hamiltonian ME's for localized conduction-band states (in eV).
Negative corrections indicate a downward shift of the conduction bands.}
\begin{tabular}{p{4.3cm}p{1.4cm}p{1.4cm}p{1.4cm}p{1.4cm}p{0.01cm}}
  \hline\hline
Cluster &[C$_{56}$] &[C$_{110}$]&[C$_{184}$]&[C$_{294}$] \\ \hline 
HF  &14.75&14.76&&\\ \hline
 &&$\Delta$H$_{nn}$($\textbf{0}$)&& \\ 
\emph{Static correlation hole} &&&&&\\
Intra-bond relaxation& --0.27&--0.28&--0.28&--0.28&\\
NN shell relaxation&--1.01 &--1.02&--1.03&--1.03\\
2nd NN shell  &&--0.65&--0.66&--0.67\\
3rd NN shell  &&&--0.51&--0.50\\
4th NN shell   &&&&--0.33\\
\emph{Long-range polarization} &--2.65&--1.81&--1.35&--1.08\\
\emph{Quantum fluctuations} &+0.30&+0.30&+0.30& +0.30\\
Total correction&--3.63&--3.46&--3.53&{\bf --3.59} \\ \hline
 \hline 
 \end{tabular} 
\label{diagonalelectron}
\end {table}

One important detail here is that the CISD treatment is not size consistent.\cite{Helgaker} Test calculations on hydrocarbon molecules indicate differences of 0.35 to 0.8 eV (from C$_2$H$_6$ to the largest molecules considered) between the correlation-induced corrections to the ionization potentials obtained by using a size consistent CC treatment and those provided by multireference CISD (MRCISD).\cite{Stoll_diamond} For example, the result from CC calculations, with single and double excitations and a perturbative treatment of triple excitations (CCSD(T)), for the combined effect of the static correlation hole and LGSC associated with intrabond (at reference C--C bond) and NN contributions is --1.89 eV for C$_2$H$_6$, as compared to a value of --1.53 eV by MRCISD.\cite{Stoll_diamond} We note that if the latter MRCISD calculation is carried out within the FLHA, the resulting MRCISD-FLHA value amounts to --1.17 eV, which indicates that the effect of the FLHA is in the range of 0.35 eV.
While for single-particle excitations, describing the relaxation and polarization cloud, size consistency effects are less important, they become sizable when dealing with two-particle excitations that describe simultaneous correlations in an
increasing number of C--C bonds, see, e.g., Refs.~\onlinecite{Gr1, Gr2, Albrecht}.
To prevent size consistency errors, Gr\"afenstein \emph{et al.}~\cite{Gr1, Gr2} and Albrecht
\emph{et al.}~\cite{Albrecht} used for the computation of correlation effects on diamond's valence bands an incremental scheme~\cite{Stoll2} where the single-particle excitations were described variationally by MRCI with singles
(MRCIS) while the two-particle excitations were treated by quasi-degenerate variational perturbation theory (QDVPT).
We extracted from Table 1 of Ref.~\onlinecite{Albrecht} the combined intrabond and NN bond 
contributions to the correlation-induced correction to the valence-band diagonal ME's.
That is obtained by summing up the relevant intrabond, 1-bond, 2-bond, and 3-bond increments computed in Ref.~\onlinecite{Albrecht} by MRCIS+QDVPT calculations on C$_8$H$_{18}$ units without the FLHA.
It amounts to --2.37 eV, 0.72 eV lower than our present value of --1.66 eV from ROHF+CISD (FLHA) calculations on the [C$_{56}$] cluster.
This 0.72 eV difference is consistent with the effect evidenced in the CCSD(T) study mentioned above and again related to size-consistency errors showing up in the present CISD treatment and to smaller extent to the FLHA. Note that the HF diagonal ME obtained by Albrecht \emph{et al.}~\cite{Albrecht} on a C$_8$H$_{18}$ molecule, --18.51 eV, compares well with the value we compute here, --18.78 eV.
\paragraph{Long-range polarization effects.}
The LR part of the polarization cloud is treated in a continuum approximation, see, e.g., Refs.~\onlinecite{Hozoi, AlexJCP, AlexPRB}. It contains polarization effects beyond the neighborhood explicitly described by \emph{ab initio} methods. 
We stress that the continuum dielectric model is well suited
for evaluating LR polarization effects beyond a large enough $R$ because the interaction of the extra charge with the surroundings beyond $R$ has predominantly electrostatic character. 
It is therefore reasonable to approximate the farther crystalline surroundings as a polarizable continuum.

The classical polarization energy of a dielectric medium caused by a charge $e$ outside a sphere of radius $R$ around the charge is given by~\cite{Fuldebook}
\begin{eqnarray}
\triangle E (R)=\frac{1}{2}\int{\bf P}.{\bf E} d^{3}{\bf r}=-\frac{\epsilon_0-1}{2\epsilon_{0}}\frac{e^{2}}{R},
\nonumber
\end{eqnarray}
where ${\bf P}$, ${\bf E}$,
and $\epsilon_{0}$ are the macroscopic polarization, unscreened electric field generated by the extra particle, and the dielectric constant of the medium, respectively. $R$ is referred to as effective correlation length~\cite{Fuldebook} and fixed to be a cutoff radius at which the dielectric function $\epsilon_0(\mathbf{r})$ reaches its asymptotic value  $\epsilon_0$. Screening due to correlation effects is fully developed at this length.

 A value $\epsilon_0$=5.7 is employed for diamond's  static dielectric constant.~\cite{AshcroftMermin}
For each cluster, the cutoff radius $R$ is determined as follows.
Each $R$ is taken as the average of the radii of two concentric consecutive ``coordination spheres'' of C--C bonds,
e.g., of the 4th and 5th coordination spheres around the reference bond in the case of the largest cluster ([C$_{294}$]) considered here. The volume and further the radius of a particular coordination sphere are obtained using 
the number of bonds contained within that sphere and the average volume per bond in diamond. The radii $R$ for the clusters [C$_{56}$], [C$_{110}$], [C$_{184}$], and [C$_{294}$] are then
2.237, 3.285, 4.396, and 5.478 \AA $\,$, respectively. For the [C$_{294}$] cluster, for example, the LR polarization correction for the valence-band hole and conduction-band electron states is found to be --1.08 eV (see Tables \ref{diagonalhole} and \ref{diagonalelectron}). The assumption hereby is that for sufficiently large values of $R$ the polarization energy is the same for both an extra hole or extra electron, except for a change of sign. 
The precise value of $R$ depends on the properties of the medium.

We assess the quality of the dielectric continuum model for diamond by comparing, for example, the \emph{ab initio} results of --0.38 eV (hole states) and --0.33 eV (electron-addition states) for relaxation and polarization effect at the shell of 4th NN C--C bonds with the estimate obtained within the continuum approximation. In the latter case, the 4th shell of C--C bonds around the reference bond hosting the extra charge is responsible for
a correction to the diagonal matrix elements of~\cite{Hozoi, AlexJCP}
\begin{equation}
\delta E_4=S_4\frac{\epsilon_0-1}{2\epsilon_{0}}e^{2}\Big(\frac{1}{R_1}-\frac{1}{R_2}\Big)\nonumber,
\end{equation}
where $R_1$= 3.866 \AA \ and $R_2$=4.926 \AA \ are the radii of the two concentric 4th and 5th coordination spheres of C--C bonds, calculated as explained above. 
The factor $S_4$ is a scaling factor representing the ratio between the density 
of C--C bonds in the spherical shell enclosed between $R_1$ and $R_2$ and the density of C--C bonds in the unit cell (see also Refs.~\onlinecite{Hozoi} and ~\onlinecite{AlexJCP}). It is equal to 1.29. The correction $\delta E_4$ is then --0.43 eV, 11\% (hole states) to 24\% (electron-addition states) larger than the \emph{ab initio} values. 
This deviation of the continuum estimate from the \emph{ab initio} values for $R$'s in this range is not surprising.  As pointed out above, the LR polarization contributions may be taken beyond a certain $R$ to be symmetric because of the equal absolute values of the polarizing charges. This is not true  for SR relaxation and polarization contributions since at shorter distances the associated charge distribution and polarization can not be approximated by those of a continuum. This situation is reflected by the different \emph{ab initio} contributions for the hole and electron-addition states due to the 4th NN bonds.
Hence, the polarization contribution of --0.43 eV deduced from the continuum model is somewhat less accurate. 
Yet its deviation from the \emph{ab initio} values is sufficiently small to conclude that in diamond the radius $R$ associated with the shell of 4th NN C--C bonds is large enough for the continuum model to yield polarization contributions of good accuracy {\it beyond} that $R$. 

We note that the LR part of the polarization cloud is well described in the $GW$ approximation. Hence, it would be pertinent to combine a treatment of the SR part of the correlation hole by QC methods with a $GW$ or random phase approximation (RPA) approach for the LR part of the polarization cloud. An attempt was made in this direction in Ref.~\onlinecite{FuldeHorsch}, yet a practical implementation turned out to be too difficult to achieve. The difficulty arises from the fact that the RPA requires working in momentum, $\mathbf{q}$-space while a short-range QC correlation treatment has to be done in real, $\mathbf{r}$-space. Relations between the $GW$ approximation and the QP approach based on QC methods are discussed in Ref.~\onlinecite{Fuldebook} (Chpt. 9) and in Ref.~\onlinecite{Fuldebook2} (see, e.g., Chpts.~4.4 and 7.3.2).

\subsection{Correlation corrections to off-diagonal ME's}\label{nondiagH}
In this section, the correlation-induced corrections to off-diagonal ME's of the 
Hamiltonian $H^{\rm eff}$ are analyzed (see Eqs.~(\ref{epsiloncorr2}) and~(\ref{Heff})).
Such corrections modify both the dispersions and the overall widths of the bands.
To this end, additional ROHF optimizations and further CISD calculations are carried out for $(N\!\pm\!1)$ states defined in terms of WO's centered on different bonds in the lattice. Renormalized ME's are then computed in terms of pairs of such electron-removal or electron-addition states, e.g., for the CISD WF's
$|\Psi^{N-1}_{\mathbf{R}_Im\sigma}\rangle$ and $|\Psi^{N-1}_{\mathbf{0}m'\sigma}\rangle$ or
$|\Psi^{N+1}_{\mathbf{R}_In\sigma}\rangle$ and $|\Psi^{N+1}_{\mathbf{0}n'\sigma}\rangle$.
As in the case of diagonal ME's, the ROHF and CISD calculations are performed within the FLHA,  yielding the $(N\!\pm\!1)$ ROHF-FLHA and CISD-FLHA states, respectively, see Sect.~\ref{diag_correlH}, Appendix~\ref{AppLGSC}, and earlier work.~\cite{Hozoi, AlexJCP,AlexPRB, interface_new}
The spin dependence of the correlation hole, see Sect.~\ref{diag_correlH}, is accounted for
just for the $(N\!-\!1)$ valence-band states.

For each ``pair'' of $(N\!\pm\!1)$ ROHF-FLHA states
$|\tilde{\Phi}^{N\pm1}_{\mathbf{R}_Ip\sigma}\rangle$ and $|\tilde{\Phi}^{N\pm1}_{\mathbf{0}p'\sigma}\rangle$,
or CISD-FLHA states $|\Psi^{N\pm1}_{\mathbf{R}_Ip\sigma}\rangle$ and $|\Psi^{N\pm1}_{\mathbf{0}p'\sigma}\rangle$,
two different, mutually non-orthogonal sets of orbitals are employed because for
each particular $(N\!\pm\!1)$ $\mathbf{R}_I$ configuration a separate ROHF optimization is carried out.
The real-space Hamiltonian and overlap ME's between such WF's are calculated by applying a nonunitary
transformation of the mutually non-orthogonal sets of orbitals to biorthogonal sets~\cite{Mitrushchenkov}
and can be further used to diagonalize the $\mathbf{k}$-matrix in Eq.~(\ref{epsiloncorr2}) to finally obtain the
QP energy bands $\epsilon_{\mathbf{k}\mu\sigma}$ and $\epsilon_{\mathbf{k}\nu\sigma}$
in Eq.~(\ref{epsiloncorr}) (see also Sect.~\ref{Sum}).
Alternatively, those ME's can be employed for deriving a set of effective hopping integrals $t_{mm'}(\mathbf{R}_{I})$ and $t_{nn'}(\mathbf{R}_{I})$ for an orthogonal tight-binding-like formulation. 
For mutually non-orthogonal ($N\!-\!1\!$) states with (slightly) different binding energies,
i.e., $S_{\mathbf{R}_I,mm'}\neq 0$ and $H_{\mathbf{0},m'm'}$ $\neq$ $H_{\mathbf{R}_I,mm}$,
the hopping integrals read~\cite{AlexJCP, AlexPRB} 
 \begin{eqnarray}
 \nonumber
 t_{mm'}(\mathbf{R}_I) &=&\frac{1}{1-S^{2}_{\mathbf{R}_I,mm'}}\\
&\times&\Bigg\{H_{\mathbf{R}_I,mm'}-
S_{\mathbf{R}_I,mm'}\bigg (\frac{H_{\mathbf{0},m'm'}+H_{\mathbf{R}_I,mm}}{2}\bigg)\Bigg\}
 \nonumber \\
 &=& \frac{1}{2}\Bigg\{\triangle E^{2}-\frac{(H_{\mathbf{R}_I,mm}-H_{\mathbf{0},m'm'})^{2}}{1-S^{2}_{\mathbf{R}_I,mm'}}\Bigg\}^{\frac{1}{2}}.
\end{eqnarray}
Analogous equations hold for $t_{nn'}(\mathbf{R}_{I})$.
$\triangle E$ is the energy separation between the two eigenstates in the $2\!\times\!2$
non-orthogonal CI (NOCI) secular problem (see, e.g., Ref.~\onlinecite{Broer}).
If $S_{\mathbf{R}_I,mm'}$=0, the expression for the hopping term reduces to
\begin{eqnarray}\label{nonort_TB}
t_{mm'}(\mathbf{R}_I)&=&H_{\mathbf{R}_I,mm'} \nonumber \\ 
&=&\frac{1}{2}\Bigg\{\triangle E^{2}-(H_{\mathbf{R}_I,mm}-H_{\mathbf{0},m'm'})^{2}\Bigg\}^{\frac{1}{2}} \,.
\end{eqnarray}

The correlation-induced corrections to the NN and 2nd NN HF ME's are obtained by calculations
on clusters of either 68 or 80 C sites.
The [C$_{68}$] cluster has an active region [C$_{11}$] that includes two adjacent reference
bonds and the eight NN C--C bonds.
There are two varieties of [C$_{80}$] clusters, [C$_{80}^{\rm cis}$] and [C$_{80}^{\rm trans}$], accounting for two possible conformations of the two reference 2nd NN bonds with respect to each other.
Each has an active region [C$_{14}$] that incorporates, in addition to the two reference C--C bonds, eleven NN bonds.
The buffer regions are always designed such that include two other coordination shells of C--C bonds around the central active regions. 
For each pair of $(N\!\pm\!1)$ states defining a particular ME, relaxation, polarization, and further correlation effects are accounted for in ROHF-FLHA and ROHF+CISD (FLHA) computations up to NN bonds around the two $(N\!\pm\!1)$ reference bonds. It turns out that the contributions from the 2nd NN bonds to the valence-band correlation-induced corrections are already smaller by a factor of 6 for the NN off-diagonal ME's.\cite{Gr1, Gr2}

For the 2nd NN ME's (and the [C$_{80}$] clusters), in order to make the CISD treatment feasible, the calculations are performed with more compact double-$\zeta$ basis sets augmented with a single $d$ function~\cite{Dunning, Birken2} (see Appendix \ref{technicaldetails}).
The effect of reducing the basis sets from triple-$\zeta$ to double-$\zeta$ on the HF hopping integrals turns out to be negligible for the valence-band states, 2 to 3 meV.
For the conduction-band HF ME's, the basis set effects are somewhat increasing, to about 5 meV for the 2nd NN \emph{trans} hopping, 50 meV for the 2nd NN $cis$ hopping, and 30 meV for the NN hopping integral.
Basis set effects in the same range were found in the CISD calculations for the NN ME's,
again more consistent for the more extended antibonding WO's.
\footnote{
Although CISD calculations for the 2nd NN hoppings are not feasible with the triple-$\zeta$ set,
the anticipated effects are most probably smaller than size-consistency errors in CISD.}
The values provided for the renormalized 2nd NN valence- and conduction-band hopping integrals in Table~\ref{Hop_renormal} are {\it a posteriori} corrected for such  basis set effects.

%%%%%%%%%%%%%%%%
%%% TABLE IV %%%
%%%%%%%%%%%%%%%%
\begin {table}[htpb]
 \centering 
\caption{NN and 2nd NN hoppings for valence-band (t$_{mm'}(\mathbf{R}_I)$) and conduction-band (t$_{nn'}(\mathbf{R}_I)$) states. Bare hopping integrals in terms of projected HF WO's and renormalized results from ROHF+CISD (FLHA) calculations are given (in eV).}
  \begin{tabular}{p{2.8cm}p{1.6cm}p{1.6cm}p{1.6cm}p{1.6cm}}
  \hline\hline
&\multicolumn{2}{c}{t$_{mm'}(\mathbf{R}_I)$}&\multicolumn{2}{c}{t$_{nn'}(\mathbf{R}_I)$} \\ 
&Bare &Renorm. &Bare &Renorm.\\ \hline
NN & 2.701&2.340&0.242&0.216\\
2nd NN {\it trans} &0.882 &0.829 &0.692 &0.731\\
2nd NN {\it cis}   &0.486 &0.477 &0.408 &0.406 \\
\hline \hline
\end{tabular} 
\label{Hop_renormal}
\end {table}
Both bare and renormalized hopping integrals are listed in Table~\ref{Hop_renormal}.
For NN hoppings, the renormalized ME's of Table~\ref{Hop_renormal} are reduced by about 10\% from the bare values. 
For 2nd neighbor hopping integrals, most of those effective ME's are again reduced when accounting for correlations, by up to 6\%.
One exception is the 2nd NN conduction-band {\it trans} hopping ME's, that is somewhat
enhanced by correlations.

Cluster size effects are not explicitly studied here for the off-diagonal ME's but
may affect to some degree the size of these integrals.
As pointed out by Birkenheuer \emph{et al.}~\cite{Birken2}, due to the larger extent of the
conduction-band WO's, the correlation-induced corrections to the off-diagonal conduction-band ME's might be of LR nature. 

Yet the most important errors appear to arise here from size-consistency effects in the CISD treatment.
Test CCSD(T) calculations on hydrogen-terminated carbon clusters show that the correlation-induced corrections
to the valence-band NN off-diagonal ME's increase by a factor of 1.6 as compared to MRCISD results obtained when relaxing the FLHA.~\cite{Stoll_diamond}
An enhancement by factors of up to 4 is additionally obtained for the 2nd NN off-diagonal ME's.~\cite{Stoll_diamond}
The effect of such errors on the QP band structure is further discussed in the next section.
\subsection{Quasiparticle energy bands}\label{Sum}
%
%FIGURE 7%%%%%%%%%%%%%%%%%%%%%
%
\begin{figure}[htbp]
\begin{center}
\caption{\label{hfcorrband} Hartree-Fock (dashed lines) and renormalized (solid lines) band structure of diamond along different symmetry lines, as obtained from embedded cluster calculations.}
\includegraphics[width=8.5cm]{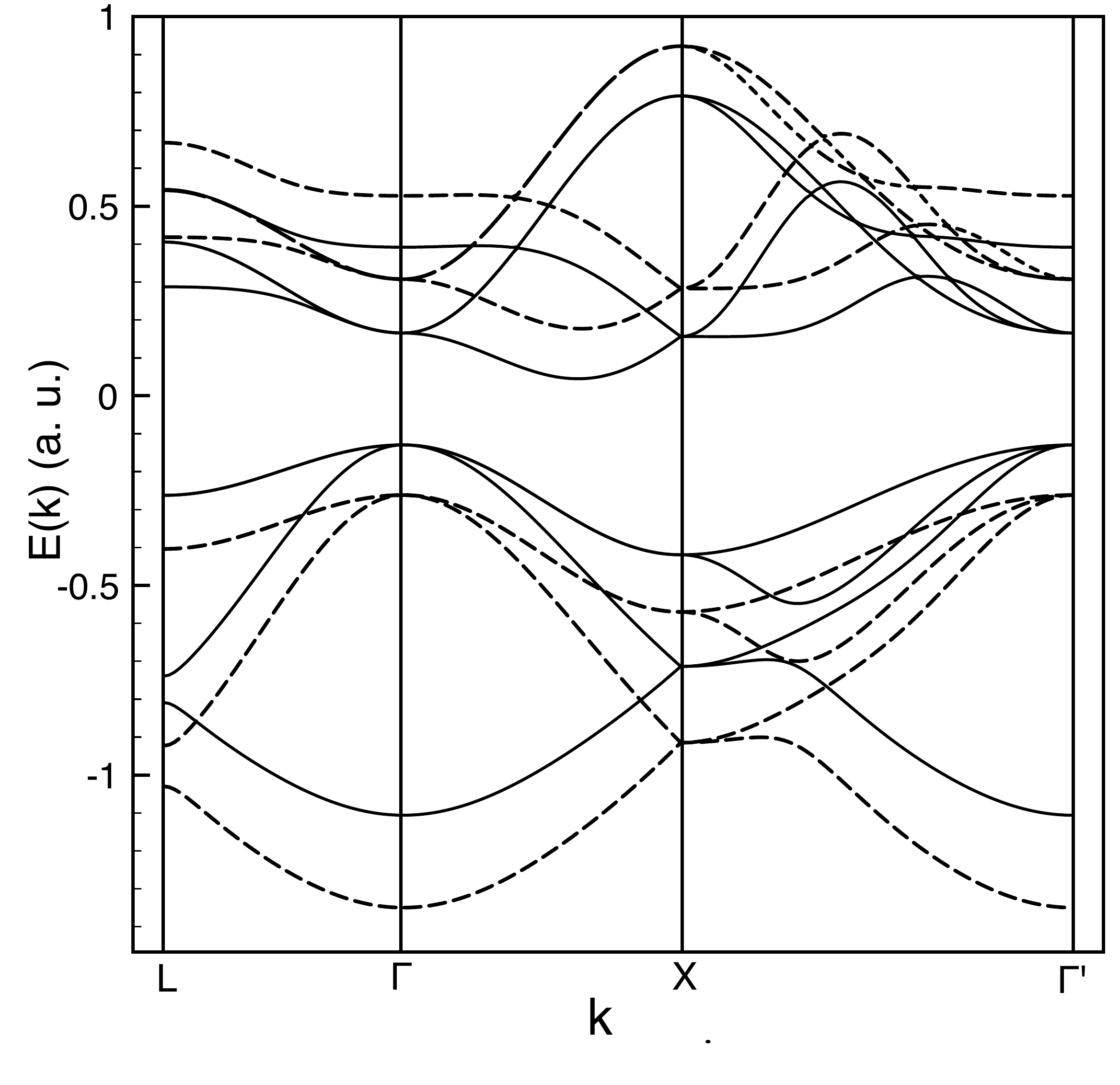}\\
\includegraphics[width=7.5cm]{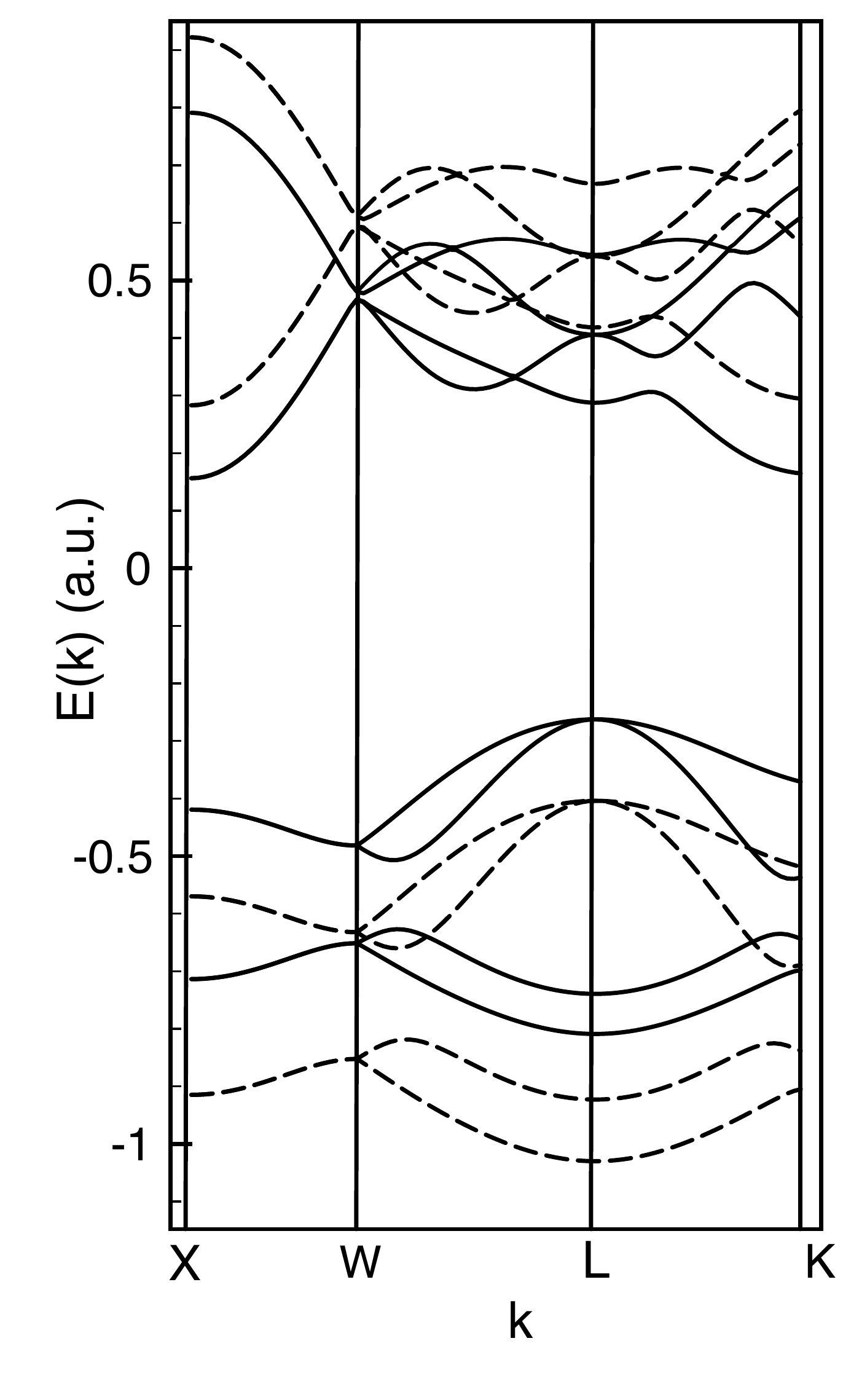}
\end{center}
\end{figure}
We first discuss the overall shift of the valence and conduction bands due to 
correlation-induced corrections to the diagonal ME's.
Summing up relaxation and polarization effects at nearby bonds up to the 4th coordination sphere around the WO with the extra particle,
the upwards correlation-induced shift of the valence bands amounts to about 4.5 eV
(see last column in Table~\ref{diagonalhole}).
This value also incorporates the estimate for the spin dependence of the correlation energy.
The correlation-induced downwards shift of the center of gravity of the four lowest-lying conduction
bands amounts to 2.8 eV (see last column in Table~\ref{diagonalelectron}).
LR polarization beyond the 4th coordination sphere further enhances those shifts,
by nearly 1.1 eV for each group of bands. Quantum fluctuations, on the other hand, more specifically the LGSC, yield opposite shifts, downwards for valence bands and upwards for conduction bands.
The net effect is a reduction of the indirect and direct HF band gaps by about 8.25 eV,
bringing the renormalized gaps to 4.32 and 6.16 eV, respectively (see top of Table \ref{final}).
%%%%%%%%%%%%%%%
%%% TABLE V %%%
%%%%%%%%%%%%%%%
\begin {table}[htp]
 \centering 
\caption{Indirect and direct gaps plus widths of the valence (VBW) and conduction (CBW) bands
at different levels of approximation. All values are in eV.
Corrections to the diagonal ($\Delta_{\rm diag}$) and off-diagonal ($\Delta_{\rm offdiag}$) ME's are reported separately. Results from \emph{G$^{0}W^{0}$} and QPsc\textit{GW} calculations from Ref.~\onlinecite{GWcalc} are also listed for comparison (see text).}
\begin{tabular}{p{3.4cm}p{1.7cm}p{1.7cm}p{1.6cm}p{1.6cm}}
  \hline\hline 
Gaps/Widths    &Indirect $\Gamma_{25'}^{v}$-$\Delta_1$&Direct $\Gamma_{25'}^{v}$-$\Gamma_{15}^{c}$&VBW&CBW\\ \hline
HF  &12.58&14.42&29.43&17.18\\
$+\Delta_{\rm diag}$&4.32&6.16&& \\
$+\Delta_{\rm offdiag}$&&&26.58&17.17\\
$+\Delta_{\rm offdiag (CC)}$&&&22.90&\\
$+\Delta_{\rm diag+offdiag}$&5.36&6.95&& \\
+CC$_{\rm VB\,correction}$ &6.08&7.67&& \\
\emph{G$^{0}W^{0}$}$^{a}$&5.73&7.39&22.20& \\
QPsc\textit{GW}$^{a}$&6.10&8.18&23.07& \\
Experiment &5.48$^{b}$&7.3$^{c}$&23.0$\pm$0.2${^d}$&
\\
 \hline \hline
 \end{tabular} 
\label{final}

$^{a}$Ref.~\onlinecite{GWcalc}
$^{b}$Ref.~\onlinecite{Clark}
$^{c}$Ref.~\onlinecite{Philipp}
$^{d}$Ref.~\onlinecite{experBWdiamond}

\end {table}

Correlation-induced corrections to the off-diagonal ME's and consequently to the widths of the bands are responsible for additional modifications of the gaps. Both the top of the valence bands and the bottom of the conduction bands, for example, are shifted downwards by different amounts due to renormalization of the off-diagonal ME's, which indeed affects the indirect and direct gaps (see Table \ref{final}).

Bare and renormalized band structures are shown in Fig.~\ref{hfcorrband}.
Off-diagonal ME's up to the 3rd neighbor are used in diagonalizing the $H(\mathbf{k})$ matrix
(see also the discussion in Sect.~\ref{hffromcluster}).
For the renormalized band structure, correlation-induced corrections are included only for
the 1st (NN) and 2nd neighbor off-diagonal ME's.
CISD and NOCI calculations for the 3rd NN off-diagonal ME's are computationally too expensive for the type of clusters, up to 184 C sites (see Sect.~\ref{hffromcluster}), and embedding scheme employed here. 
Earlier studies on hydrogen-terminated carbon clusters indicate, however, rather 
small corrections for the 3rd NN ME's, 3 to 20 meV, due to correlation effects at the NN bonds around the two reference $(N\!\pm\!1)$ bonds.\cite{Gr1}

Overall upwards/downwards shifts of 3.60/3.86 eV are now found at the $\Gamma$ point for the renormalized upper-valence/lowest-lying conduction bands. The conduction-band bottom ($\Delta_{\rm min}$) is also shifted downwards by about 3.59 eV.
Due to such shifts applied to the \textsc{crystal} HF bands, the indirect and direct band gaps are reduced by 7.22 and 7.47 eV, respectively, to values of 5.36 and 6.95 eV (see Table~\ref{final}), close to experimental estimates of 5.5~\cite{Clark} and 7.3 eV.\cite{Philipp}
The width of the valence band is also significantly modified. For example, a reduction from 29.43 to 26.58 eV is found at the ROHF+CISD (FLHA) level for the valence-band width (VBW), see Table~\ref{final}.
The evolution of the latter due to renormalization of the hoppings can be easily tracked
down by using the tight-binding relation for, e.g., $\mathbf{k}$=0:
$\Delta_{\rm { VBW}}$$(\Gamma^{v}_{25'}$-$\Gamma^{v}_1)$=8$t^{\rm NN}_{mm'}(\mathbf{R}_1)$+16t$^{\rm 2nd NN, cis}_{mm'}(\mathbf{R}_2)$. 
The results listed in Table~\ref{Hop_renormal} suggest that the largest corrections come from renormalization of the NN hopping integrals.
We note that accounting for correlation effects at the 2nd neighboring bonds on the NN off-diagonal  ME's leads to a small change in $\Delta_{\rm {VBW}}$, of about 0.1 eV. The same conclusion was reached in earlier MRCIS+QDVPT studies of the valence-band states by Gr\"afenstein \emph{et al.}\cite{Gr1, Gr2}, using hydrocarbon models.

The ROHF+CISD (FLHA) VBW still is too large as compared to experiments,\cite{experBWdiamond} by about 3 eV (see Table~\ref{final}). Relaxing the FLHA constraint in test MRCISD calculations on hydrocarbons~\cite{Stoll_diamond} leads to a reduction of the VBW by less than 1 eV.
As discussed in Sect.~\ref{nondiagH}, numerical tests comparing the outcome of CCSD(T) and MRCISD (with relaxed FLHA) calculations on hydrogen-terminated carbon clusters show that the correlation-induced modifications of the 1st (NN) and 2nd neighbor off-diagonal ME's are enhanced in the CCSD(T) treatment, i.e., the reduction of the hopping ME's is stronger with the size-consistent CC approach.~\cite{Stoll_diamond}
Porting those differences between the CCSD(T) and MRCISD results for the molecular-like hydrogen-terminated carbon units to our embedded-cluster data, gives rise to a further reduction of the VBW to 22.90 eV, see Table~\ref{final}. Additionally, the indirect and direct gaps are now changing to 6.08 and 7.67 eV, respectively, see the lower lines in Table~\ref{final}. 

Other features of the valence bands that are affected by correlations are, for example, the coupling strength of the avoided crossing along the X--$\Gamma'$ line and the shape and widths of the bands along the W--L and L--K symmetry directions (see Fig.~\ref{hfcorrband}). For the four low-lying conduction bands, besides the downwards shift,
no other pronounced correlation-induced modifications were observed at the CISD level.
However, as also discussed in Sect.~\ref{nondiagH}, further studies are needed to assess longer-range correlation effects for the off-diagonal conduction-band ME's as well as the effect of size-consistency errors.
We note that the FLHA approach is not implemented at present for a CC ansatz.
CC calculations with the present cluster-in-solid embedding scheme and an unrestricted hole approach (as for hydrocarbons) turned out to be computationally too expensive as regards the off-diagonal ME's.

It is instructive to compare our results for the band gaps and VBW with values from Table I in Ref.~\onlinecite{GWcalc} as obtained from non-self-consistent (one-shot, \emph{G$^{0}W^{0}$}) and  quasiparticle self-consistent \emph{GW} (QPsc\emph{GW}) calculations~\cite{Faleev} (see Table~\ref{final}). The self-energy $\Sigma^{0}$ is computed thereby from DFT-GGA eigenfunctions~\cite{GWcalc}. As for other $sp^{3}$
bonded materials~\cite{HL, HL2, Falco, Onida, Godby, Faleev, Schilfgaarde}, the 
fundamental gap of diamond is significantly improved in \emph{G$^{0}W^{0}$} (and in QPsc\emph{GW}) as compared to DFT-GGA~\cite{GWcalc}. This is not surprising since the \emph{GW} approximation treats explicitly two effects that are essential for the band gap -- non-local exchange and correlations.
Whereas the LR polarization cloud is described very well in \emph{GW}, the treatment of SR relaxation and polarization is much less accurate.
The relatively good agreement between \emph{GW} results and experimental estimates (see Table~\ref{final}) shows that the drawback mentioned above seem not to have a major impact for the fundamental band gaps of $sp$ covalent semiconducting materials~\cite{Schilfgaarde}. For diamond, our LHA results and the  \emph{G$^{0}W^{0}$} and QPsc\emph{GW} data from Ref.~\onlinecite{GWcalc} are comparable as regards the direct and indirect band gaps. 
As \emph{G$^{0}W^{0}$} depends on the LDA/GGA solutions, it might yield quantitatively crude QP energies if the DFT solution is a crude starting point~\cite{Schilfgaarde, Schilfgaarde2, Faleev}. To obtain reliable results for a wider range of materials, self-consistency is in principle desirable. Yet while the self-consistent  QPsc\emph{GW} diamond's VBW is indeed improved as  compared to the \emph{G$^{0}W^{0}$} result, this seems not to be the case for band gaps, in particular, the direct gap.

\section{Conclusions and outlook}~\label{Concl}
The aim of this work is to assess to which accuracy \emph{ab initio}
correlation calculations for conduction and valence bands of semiconductors and insulators can be presently  performed.
Diamond, a prototype \emph{sp} covalent insulator, is here selected for this purpose.
The calculations are based on many-body wave-function computations 
and are an alternative to band-structure studies resting on density functional theory. The quasiparticle approximation is thereby adopted, i.e., damped or one damped eigenmodes of the correlation hole of an added particle (electron or hole) are not taken into account.
The starting point in the correlation treatment is the Hartree-Fock band structure,
obtained by using the \textsc{crystal} package for periodic-structure calculations. 
It is shown that with a properly constructed embedding scheme and large enough clusters,
reference energy bands nearly identical with the initial Hartree-Fock bands from the periodic
calculation can be obtained. Correlation-induced corrections to the HF bands have been next computed by means of an effective local Hamiltonian approach.
For the effect of correlations on the energy bands a distinction is made between the short-range modifications in the surrounding of the added particle (short-range correlations) and the long-range part of the polarization cloud generated by that particle.
The short-range part is treated to the highest accuracy we can presently achieve by considering an embedded cluster with the added particle kept frozen in a Wannier orbital localized on a reference site/bond in the cluster. Various clusters containing up to 294 carbon atoms are utilized, embedded in effective one-electron potentials that are derived explicitly from a prior Hartree-Fock periodic calculation for the neutral crystal. First the static part of the correlation holes is computed by determining the changes in the electronic density around the added particle (charge relaxation and polarization).
This is achieved by means of additional restricted open-shell Hartree-Fock optimizations where contributions
from orbitals localized at up to the 4th NN C--C bonds around the extra particle are explicitly computed. 
Quantum fluctuations, more specifically loss of ground-state correlations, are then calculated by configuration-interaction schemes with one- and two-particle excitations. The resulting many-body wave functions describing the short-range part of the correlation hole constitute the space within which
the matrix elements of the effective local Hamiltonian are computed.
The correlated energy bands within the quasiparticle approximation are 
expressed in terms of those matrix elements. 
The contribution of the long-range part of the polarization cloud, extending beyond the region within which  correlations are explicitly treated, is handled by a dielectric continuum approximation. 
It is shown that this approximation is well suited for the evaluation of long-range polarization effects beyond a sphere that encompasses all C--C bonds up to 4th NN's around the bond hosting the extra charge. 

The net outcome of the present and earlier quantum-chemistry work~\cite{Hozoi,AlexJCP,AlexPRB,interface_new,QP_bands_cuprates}
is demonstrating that wave-function energy band calculations of good quality can be performed not only for valence but also for conduction bands of semicondutors and insulators.
One virtue of wave-function-based approaches such as the LHA is that they rely on 
well-defined and controllable approximations for the description of the short-range part of the correlation hole in direct, $\mathbf{r}$ space.
One shortcoming, for the time being, is that the quantum fluctuations are treated by CI calculations that are not size consistent. For diamond, in particular, this gives rise to an understimation of the correlation-induced reduction of the valence-band width as compared to experimental estimates. Therefore, it is  desirable to establish a size-consistent approach -- employing, e.g., a
CC or coupled electron pair approximation (CEPA) wave function ansatz -- for simultaneously evaluating relaxation, polarization, and quantum fluctuations effects. This is possible, yet requires further programming efforts.
When this is done one will have a powerful and highly accurate tool available to compute
energy bands for semiconductors and insulators.
\subsection*{Acknowledgements}A.~S. acknowledges useful discussions with V.~Katukuri.
%

%%%%%%APPENDICES%%%%%%%%%%%%%%%%%%%%%%%%%%%%%%
\appendix
\section{Computational details}\label{technicaldetails}
Diamond is a conventional $sp$ insulator with face-centered cubic structure and Fd$\bar{3}$m space-group symmetry. The local point-group symmetry for each C atom is T$_d$, with a four-site NN coordination, and covalent C--C bonds. Its lattice constant is 3.567 \AA~\cite{laticediam}, with an interatomic C--C distance of 1.545 \AA.

In the crystal and cluster calculations, we employed all-electron GTO basis sets of triple-$\zeta$ quality augmented with one single $d$ polarization function \cite{Dunning, Birken2}. For some 2nd NN off-diagonal ME's of H$^{\rm eff}$, GTO basis sets of double-$\zeta$ quality and a single $d$ function \cite{Dunning, Birken2} were also used to enable their calculation. 
Both basis sets were optimized for diamond~\cite{Birken2}.
For the \textsc{crystal}\cite{CRYSTAL1} periodic HF calculations, a Monkhorst-Pack grid \cite{Monkhorst} with a shrinking factor IS=14 is used. The thresholds for the integral calculations were chosen tight, (TOLINTEG parameters of  10, 10, 10, 10,  20, see Ref.~\onlinecite{CRYSTAL1}) and the total HF energy is converged with an accuracy of 10$^{-11}$ a.~u.~
To obtain the valence- and conduction-band WO's, we utilized the Wannier-Boys orbital localization module~\cite{Zikovich} in \textsc{crystal}. Projected WO's for the finite clusters $\mathcal{C}$ as well as the matrix representation of the crystal
Fock operator $F^{\rm cryst}$ in the set of GTO basis functions within $\mathcal{C}$ are calculated with a new, improved interface program \cite{interface_new} stemming from the original \textsc{crystal-molpro} interface \cite{interface}. The interface program yields in addition the virtual PAO's.
The matrix representations of the density $P^{\mathcal{C}}$ and Fock $F[P^{\mathcal{C}}]$
operators of $\mathcal{C}$ in the GTO basis set as well as the embedding potential $V^{\rm emb}$ are calculated with the \textsc{molpro} \cite{molpro} suite of programs. The correlation calculations are also carried out with \textsc{molpro}.
\section{Projected WO's and $V^{\rm emb}$}\label{AppA}

As mentioned in Sect.~\ref{HF_bands}, we used Mulliken population data to estimate the spatial extent of the crystal WO's.\cite{Zikovich}. 
The analysis shows that for the antibonding WO's, the atomic population at each of the two atoms defining a C--C bond is the largest one, 0.44 e per C site.
The overall population associated with the NN's of the C pair is 0.1 e. Each of the four bonding WO's has a slightly larger atomic population of 0.49 e per C site, indicating a faster decay with the distance from the bond. Tiny overall populations are associated with the nearest and 2nd nearest bonds.  

The original WO's $\vert w_{p}(\mathbf{R}_I)\rangle$ centered in $\mathcal{C}$ are projected onto the set of GTO basis functions $\{\alpha\}$ assigned to the cluster's sites~\cite {Birken2, Hozoi, AlexJCP, AlexPRB}
\begin{eqnarray}
\vert w'_{p}(\mathbf{R}_I)\rangle=\sum_{\alpha, \beta \in \mathcal{C}}\vert \beta\rangle S^{-1}_{\beta\alpha}\langle\alpha\vert w_{p}(\mathbf{R}_I)\rangle \ p = m\sigma, n\sigma \nonumber
\end{eqnarray}
with $S_{\beta\alpha}$ being the corresponding overlap matrix.  
The longer-range tails of $\vert w_{p}(\mathbf{R}_I)\rangle$ involving GTO's at bonds outside $\mathcal{C}$ and orthogonal to the GTO's in $\mathcal{C}$ are thus cut off. 
If the cluster $\mathcal{C}$ is large enough, the WO's $\vert w'_{p}(\mathbf{R}_I)\rangle$ centered in $\mathcal{C_A}$ are practically identical to the original WO's.
Given the way the embedding potential is constructed (see main text and references therein), 
the asymmetric representation in the $\{\alpha\}$ basis  of WO's $\vert w'_{p}(\mathbf{R}_I)\rangle$ near the edges of $\mathcal{C}$,
i.e., within the buffer $\mathcal{C_B}$ region, does not affect the accuracy of the post-HF
cluster calculations.

The projected WO's undergo groupwise orthonormalization.~\cite{Birken2, Hozoi, AlexJCP, AlexPRB} The real-space ($N\pm1$) states are then expressed in terms of the resulting orthonormalized WO's $w''_{p}(\mathbf{R}_I)$. The variational orbital space in the correlation calculations includes the active WO's $w''_{p}(\mathbf{R}_I)$ plus the set of PAO's within $\mathcal{C_A}$ as virtual orbitals. These virtual PAO's are generated from the GTO's basis functions assigned to $\mathcal{C_A}$ sites by projecting out the occupied WO's $w''_{m}(\mathbf{R}_I)$ and lowest-lying unoccupied WO's $w''_{n}(\mathbf{R}_I)$ of $\mathcal{C}$ via a Schmidt orthogonalization.~\cite{Pulaypaos, Hampel} All those sets of orbitals are next L\"owdin orthonormalized among themselves.~\cite{Hozoi, AlexJCP, AlexPRB}. The PAO's are orthogonal to the occupied and lowest-lying unoccupied WO's in $\mathcal{C_A}$ by construction. Those in $\mathcal{C_A}$ are identical with the crystal virtual orbitals, while those in $\mathcal{C_B}$ deviate from the latter. 

The embedding potential $V^{\rm emb}$ is obtained as $V^{\rm emb}$=$F^{\rm cryst}$--$F[P^{\mathcal{C}}]$, 
where $F^{\rm cryst}$ is the crystal Fock operator
and $F[P^{\mathcal{C}}]$ the Fock operator of $\mathcal{C}$.~\cite{Birken2, Hozoi,  AlexJCP}
The density operator $P^{\mathcal{C}}$
is constructed from the projected orthonormalized occupied WO's in $\mathcal{C}$.
Since in the cluster calculations $V^{\rm emb}$ is added back to $F[P^{\mathcal{C}}]$, the correlation treatment is carried out effectively in an infinite frozen HF environment (see also Refs.~\onlinecite{AlexJCP},~\onlinecite{McWeeny}, and~\onlinecite{Werner}). 

To check that $V^{\rm emb}$ ensures consistency between the crystal and cluster HF WF's, $N$-particle HF GS calculations can be carried out for a given embedded cluster such that only the projected WO's in $\mathcal{C_A}$ are optimized. The changes in the cluster HF total energies are found to be vanishingly small, in the range of a few meV to a few tenths of eV.
This shows that $i)$ the projected WO's centered in $\mathcal{C_{A}}$ are practically identical to their original counterparts, with very tiny changes for the LR
tails and $ii)$ for each of the clusters we designed, the embedding prevents artifacts such as an artificial polarization of the charge distribution within $\mathcal{C}$. The charge distribution within $\mathcal{C_A}$ is that in the infinite system.

\section{Calculation of the quantum fluctuations -- LGSC} \label{AppLGSC}
To compute the relevant quantum fluctuations, i.e., the LGSC, we carried out CISD calculations with the reference 
ROHF-FLHA WF's $|\breve{\Phi}^{N-1}_{\mathbf{R}_Im\sigma}\rangle$ and 
$|\tilde{\Phi}^{N+1}_{\mathbf{R}_In\sigma}\rangle$. Under the FLHA-constraint, the CISD WF contains only 
singly and doubly excited configurations in which the localized orbital hosting the extra particle is always singly-occupied.   
For the ($N\!-\!1\!$) states, for example, that constraint corresponds to projecting the unconstrained CISD WF
$|\bar{\Psi}^{N-1}_{\mathbf{R}_Im\sigma}\rangle$ (from an unrestricted hole approach like that in Ref.~\onlinecite{Gr1}) onto a model space spanned by WF's composed of singly and doubly excited configurations obtained from the reference ROHF-FLHA WF $|\breve{\Phi}^{N-1}_{\mathbf{R}_Im\sigma}\rangle$. The extra hole always resides at a reference bonding WO in the center of the active [C$_8$] region of cluster [C$_{56}$].

First, a zero-order estimate of the LGSC for the ($N\!-\!1\!$) state can be obtained from the total CISD correlation energy  for the $N$-electron GS or from relevant fractions thereof. 
The CISD WF for the GS is obtained by correlating all 7 doubly occupied bonding orbitals in [C$_8$].  The resulting CISD correlation energy amounts to --7.84 eV. The electron intra-pair correlation contribution in the reference bonding orbital is missing for the ($N\!-\!1\!$) state as compared to the $N$-particle state. 
The inter-pair contribution from the correlation of that reference electron pair with electron pairs in the 6 NN bonding orbitals decreases by half in the ($N\!-\!1\!$) state because of the missing electron. The sum of the two contributions, 1.7 eV, provides a first rough estimate of the LGSC. It assumes that the orbitals of the ($N\!-\!1\!$) state are identical to those of the $N$-electron GS in the CI calculations. 

Further, constructing the CISD WF $|\Psi^{N-1}_{\mathbf{R}_Im\sigma}\rangle$ by correlating all 7 C--C bonding orbitals in [C$_8$] under the FLHA constraint, we obtain a value of 0.91 eV for the LGSC (see Table~\ref{diagonalhole}). It is calculated as the difference between the GS and ($N\!-\!1\!$) CISD-FLHA correlation energies. This finding illustrates the importance of  orbital relaxations for the  ($N\!-\!1\!$) state.
The correction to the diagonal ME's due to LGSC shifts the valence $sp$ bands downward by 0.91 eV. 

Analogously, the CISD-FLHA WF's $|\Psi^{N+1}_{\mathbf{R}_In\sigma}\rangle$ are calculated by correlating all 7 bonding orbitals in [C$_8$]. The ``frozen'' singly-occupied antibonding WO where the extra electron resides is part of the active orbital space, but no excitations out of that orbital are included.  The LGSC for the ($N\!+\!1\!$) state amounts to 0.3 eV (see Table~\ref{diagonalelectron}) and leads to an upward shift of the lower-lying conduction bands. 

Finally, we comment on the FLHA. ``Freezing'' the added particle into a WO of the $N$-particle system might not be always optimal. If FLHA is relaxed in test MRCISD calculations on hydrocarbon molecules (see Sect.~\ref{diag_correlH}), the correlation-induced corrections to the valence-band diagonal ME's are moderately enhanced, by 0.3 to 0.35 eV. 
Therefore, we do not expect significant changes in the ROHF and CISD results in Table~\ref{diagonalhole} if the FLHA is relaxed. 
%%%%%%%%%%%%%%%%%%%%%%%%%%%%%%%%%%%%%%%%%%%%%%%%%%%%%%%%%%%%%%%%%%%%%%%%%%%%%%%%%%%%%%%%%%%%%%%%%%%%%%%%%%%%%%%%%%%%%%%%
%
\section{Renormalized hopping integrals from ROHF calculations}\label{AppRenorml}
The effective hopping integrals calculated from off-diagonal ME's between the ROHF-FLHA WF's $\vert\tilde{\Phi}^{N+1}_{\mathbf{R}_In\sigma}\rangle$ and $\vert\tilde{\Phi}^{N+1}_{\mathbf{0}n'\sigma}\rangle$ or $\vert\tilde{\Phi}^{N-1}_{\mathbf{R}_Im\sigma}\rangle$ and $\vert\tilde{\Phi}^{N-1}_{\mathbf{0}m'\sigma}\rangle$ are listed in Table~\ref{Hop_renormal2}. For the ($N\!-\!1\!$) states, the effect of relaxing the orbital hosting the hole is also accounted for.
\begin {table}[htbp]
 \centering 
  \caption{
NN and 2nd NN effective hopping ME's  for valence-band (t$_{mm'}(\mathbf{R}_I)$) and conduction-band  (t$_{nn'}(\mathbf{R}_I)$) states. Bare hoppings in terms of projected HF WO's and renormalized values from ROHF calculations are given (in eV).}
  \begin{tabular}{p{2.8cm}p{1.6cm}p{1.6cm}p{1.6cm}p{1.6cm}}
  \hline\hline
&\multicolumn{2}{c}{t$_{mm'}(\mathbf{R}_I)$}&\multicolumn{2}{c}{t$_{nn'}(\mathbf{R}_I)$} \\ 
&Bare&Renorm.&Bare&Renorm.\\ \hline
NN & 2.701&2.497&0.242&0.197
\\
2nd NN \emph{trans}&0.882 &0.886&0.692&0.785\\
2nd NN \emph{cis}&0.486&0.499&0.408&0.428\\
\hline \hline
\end{tabular} 
\label{Hop_renormal2}
\end {table}
As discussed in Sect.~\ref{diag_correlH}, the ROHF-FLHA WF's describe the static relaxation and polarization cloud around each of the two reference bonds where the WO's with the extra particle (hole or electron) are centered. 
The polarizability of the nearby bonds is reduced when quantum fluctuations (i.e., LGSC) are also accounted for as in the CISD calculations. This reflects the fact that it is more difficult to polarize correlated than uncorrelated electrons in bonds. Since the static relaxation and polarization cloud is built up from uncorrelated electrons in ROHF (bare electron-hole pairs) and the restricted HF does not account for the different polarization of $\alpha$ and $\beta$ electrons in the bonds, a larger bond polarization might be obtained. Larger 
charge polarization effects may lead to either smaller or larger renormalization for the bare hopping integrals as compared 
to the CISD treatment. This is illustrated by the renormalized hopping terms in Table~\ref{Hop_renormal2} obtained from the ROHF calculations. 
One possible, straightforward way to incorporate spin polarization in the HF approach would consist in treating polarization effects around the extra hole or electron by means of unrestricted HF. 
%
%%%%%%%THE BIBLIOGRAPHY%%%%%%%%%%%%%%%%%%%%%%%

\end{document}